\documentclass[12pt]{iopart}

\usepackage{graphicx}
\usepackage{iopams}
\usepackage{dsfont}
\usepackage{fancyhdr}

\usepackage{color}
\newcommand{\grm}{\mathrm{g}}
\newcommand{\orm}{\mathrm{o}}
\newcommand{\link}[1]{{#1}}

\begin{document}
\title[Visualization of the G\"odel universe]{Visualization of the G\"odel universe}
\author{M~Buser, E~Kajari and W~P~Schleich}
\address{Institut f\"ur Quantenphysik and Center for Integrated Quantum Science and Technology (\textit{IQ}$^{ST}$), Universit\"at Ulm, Albert-Einstein-Allee 11, D-89069 Ulm, Germany}
\ead{wolfgang.schleich@uni-ulm.de}

\begin{abstract}
The standard model of modern cosmology, which is based on the Friedmann-Lema\^itre-Robertson-Walker metric, allows the definition of an absolute time. However, there exist (cosmological) models consistent with the theory of general relativity for which such a definition cannot be given since they offer the possibility of time travel. The simplest of these models is the cosmological solution discovered by Kurt G\"odel, which describes a homogeneous, rotating universe. Disregarding the paradoxes that come along with the abolishment of causality in such spacetimes, we are interested in the purely academical question how an observer would visually perceive the time travel of an object in G\"odel's universe. For this purpose, we employ the technique of ray tracing, a standard tool in computer graphics, and visualize various scenarios to bring out the optical effects experienced by an observer located in this universe. In this way, we provide a new perspective on the space-time structure of G\"odel's model. 
\end{abstract}
{\vspace*{1ex}}
Online supplementary data available from \link{stacks.iop.org/NJP/15/013063/mmedia}

\maketitle%
\tableofcontents%

\pagestyle{fancy}
\fancyhf{}
\fancyhead{}
\fancyfoot{}
\fancyhead[R]{\leftmark}
\fancyhead[L]{\thepage}
\renewcommand{\headrulewidth}{0pt}
\renewcommand{\headheight}{16pt}

\section{Introduction\label{sec:introduction}}

In the science fiction novel {\it The Time Machine} (1895) by H. G. Wells an English scientist constructs a machine which allows him to travel back and forth in time. He uses this device to visit the world far in the future but returns from his journey only a few hours after he has started it. What sounds so unproblematic in the novel is an enigma in the world of physics since the fascinating idea of time travel leads inevitably to the puzzling question: ``[...] how can we interpret the process that an agent in principle could kill his own grandfather in his (the grandfather's) childhood, thereby destroying the basic conditions for his own existence?''~\cite{Pfa81}. 

The assumption of a self-consistency principle~\cite{John90,Echeverria91}, which restricts the set of all possible initial conditions to those that lead to self-consistent solutions of the laws of physics, could avoid such logical paradoxes. However, it seems very unlikely that time travel is possible in practice. As Stephen Hawking postulated in his {\it chronology protection conjecture}: ``The laws of physics do not allow the appearance of closed time-like curves.''~\cite{Haw92}. He argued that on a submicroscopic scale time travel is prevented by the laws of nature. Beside all the physical aspects adduced in his paper, Hawking bolstered his conjecture with the simple and illustrative conclusion ``[...] we have not been invaded by hordes of tourists from the future'' yet.

The history of the scientific controversy about the possibility of time travel can be traced back to the ingenious logician Kurt G\"odel. He entertained some doubt~\cite{FefermanVolIII} about the existence of an objective time in the Newtonian sense, where time is considered as infinite layers of ``now'' coming into existence successively~\cite{Rin09}. In an essay~\cite{Goedel1949} published to honour Albert Einstein on the occasion of his 70th birthday, G\"odel adumbrated the existence of a ``cosmological solution of another kind than those known at present, to which [...] defining an absolute time is not applicable''. However, at this point he left open what he meant.

Only a few months later, in July 1949, G\"odel published the paper ``An example of a new type of cosmological solution of Einstein's field equations of gravitation''~\cite{Goe49} in which he described the space-time of a homogeneous, rotating universe. The intriguing mystery of his new model is the existence of closed time-like curves (CTCs) which make it feasible to go on a mind-boggling journey into ones own past. Travelling along such a CTC eventually leads the observer back to the point in space and time where he started from. However, the observer's own proper time elapses as usual during this trip.

CTCs in a cosmological solution are profound since they abolish causality impetuously defeated by standard physics. The reason for this breakdown of causality can be ascribed to the intrinsic rotation of the G\"odel space-time which reflects itself in a uniform gravitomagnetic field~\cite{Costa08}. As \cite{FefermanVolIII} reveals, G\"odel was very well aware of the connection between the rotation of a space-time and its temporal structure. Indeed, it was this connection which finally prompted him to target ``rotating solutions'' since ``[...] in a cosmological model, non-rotation of the major mass points is precisely the necessary and sufficient condition for there to exist a natural notion of simultaneity relative to their worldlines.'' \cite{FefermanVolIII}.

However, another aspect could also have kindled G\"odel's interest in seeking for rotating solutions \cite{Jun06}. In October 1946, George Gamow published the article ``{\it Rotating Universes?}'' \cite{Gam46} in {\it Nature} which deals with the rotation of mass aggregations such as planets, stars and galaxies. According to Gamow, ``galaxies are found in the state of more or less rapid axial rotation'' which is in contradiction to the statistical distribution of angular momentum if we regard the galaxies as condensation of the primordial matter. Driven by this contradiction, Gamow speculated in the short note \cite{Gam46} that maybe ``all matter in the visible universe is in a state of general rotation around some centre located far beyond the reach of our telescopes?''. He conjectured ``[...] that in the language of the general theory of relativity such a rotating universe can be probably represented by the group of anisotropic solutions of the fundamental equations of cosmology'' \cite{Gam46} 
unsuspecting that G\"odel a few years later would come up with exactly such a solution.

In his paper~\cite{Goe49} G\"odel admitted that his rotating universe cannot serve as a model of the universe we live in since it does not contain any redshift for distant objects accounting for an expansion as required by Edwin Hubble's law of  1929~\cite{Hub29}. Nevertheless, G\"odel's paper~\cite{Goe49} stimulated in the aftermath a lot of discussions about rotating solutions, Mach's principle~\cite{Pfister1995,Ciufolini95,Bondi1997} and CTCs. Other space-times such as the Kerr metric \cite{Ker63, Car68}, or the Gott universe of two cosmic strings \cite{Got91, Bir00} contain CTCs. Moreover, the use of a wormhole as a time machine was suggested in~\cite{Fro90}. The van Stockum space-time of a rotating dust cylinder \cite{Sto37} published 12 years before G\"odel came up with his solution already contained CTCs as well. However, these CTCs were not discovered at this time yet.

Unfortunately nature has not given us a sense to grasp the structure of a given space-time directly. Therefore, our intention is to bridge the gap between the pure mathematical description of G\"odel's universe and the wish of the reader for {\it Anschaulichkeit}. We present a series of detailed illustrations and visualizations in the spirit of \cite{Grave08,Kaj09, Fei09} to bring out the space-time structure and the compelling optical properties arising from the curvature of this specific cosmological model. Our particular interest is in the visual perception of a time travelling object that moves along a CTC in G\"odel's universe. We thereby refrain from entering the discussion on paradoxes in connection with time travel. Instead, we restrict ourselves to the very simple setting in which a sphere starts its journey near the position of an observer, then takes a single loop on a CTC before it returns to its initial position. Of course, G\"odel's universe is not the only model that could be used for such a 
study. However, our specific interest in G\"odel's model is best summarized by the concise statement of Wolfgang Rindler: ``[...] G\"odel's universe was the cleanest example, certainly the one that caught the widest attention, and possibly the first where time loops were explicitly recognized''~\cite{Rin09}.

This paper is organized as follows: In section~\ref{sec:light_propagation} we give a brief introduction to ray tracing and its adaption to curved space-time which we shall use for visualizing several scenarios in G\"odel's universe. This method requires a detailed knowledge of light propagation in the considered space-time, which also is addressed in this section. In section~\ref{sec:stationary_scenarios} we visualize objects at rest, while in section~\ref{sec:dynamic_scenarios} we present the view on freely falling objects. In section~\ref{sec:time_traveling} we discuss the visual appearance of time travelling objects, i.e. objects moving on a CTC. We conclude by summarizing our results and presenting an outlook in section~\ref{sec:summary}.

\section{Light propagation in G\"odel's universe\label{sec:light_propagation}}
\enlargethispage*{2ex}

We start by providing a brief introduction into the G\"odel universe. In order to give an overview of the causal structure of the G\"odel space-time, we discuss in section~\ref{subsec:metric} its metric and the light cone diagram. We continue by presenting in section~\ref{subsec:geodesics} the null geodesics which are of particular interest for our visualizations since they describe the paths along which light propagates. Thus they are essential for the appearance of objects to an observer. In order to complete our considerations of propagating light, we analyse in section~\ref{subsec:rayoptics} how a light front in the G\"odel space-time expands.

Null geodesics are solutions of the geodesic equation which fulfil the light-like condition $u^{\mu}u_{\mu}=0$ for the four-velocity vector $u^{\mu}\equiv\frac{\rmd x^{\mu}}{\rmd\lambda}$ where $\lambda$ represents the curve parameter along the path $x^{\mu}(\lambda)$. For a detailed derivation of the solution of the geodesic equation based on the intrinsic symmetries of the G\"odel space-time we refer to~\cite{Kaj09}. Our discussion of the null geodesics reveals the existence of an optical horizon. This phenomenon has a crucial impact on the visualizations presented in the following sections. In order to lay the foundations for these presentations, we briefly review in section~\ref{subsec:raytracing} the basic ideas behind ray tracing and show in section~\ref{subsec:goedelhorizon} first applications to the $(x,y)$-plane illustrating the critical G\"odel horizon.

\subsection{Metric, symmetries and light cones\label{subsec:metric}}

The metric of G\"odel's universe describes a homogeneous and anisotropic space-time which contains a (pressure-free) perfect fluid. According to \cite{Kaj04}, the line element $\rmd s_G^2=g_{\mu\nu}\rmd x^{\mu}\rmd x^{\nu}$ of the G\"odel space-time is given by the metric coefficients
\begin{equation}
g_{\mu\nu}(x^\sigma)\equiv
\left(\begin{array}{cccc}
c^2 & 0 & r^2\,\frac{c}{\sqrt{2}a} & 0 \\
0 & -\frac{1}{1+\left(\frac{r}{2a}\right)^2} & 0 & 0 \\
r^2\,\frac{c}{\sqrt{2}a} & 0 & -r^2\left(1-\left(\frac{r}{2a}\right)^2\right) & 0 \\
0 & 0 & 0 & -1
\end{array}
\right)
\label{eq:metric}
\end{equation}
with the speed of light $c$, the G\"odel parameter $a>0$ and the coordinates $t$, $r$, $\phi$ and $z$. Since in~(\ref{eq:metric}) the fraction $r/(2a)$ needs to be dimensionless, we assign to the parameter $a$ the unit of a length. At this point we note that for the radial coordinate $r=2a$ the coefficient in front of $\rmd\phi^2$ vanishes. This distance, which we designate as the critical G\"odel radius $r_G\equiv 2a$, is salient in the discussion of the G\"odel universe.

In the limit $a\rightarrow\infty$ the G\"odel line element $\rmd s_G^2$ converges towards the Minkowski line element $\rmd s_M^2 \equiv c^2 \rmd t^2 - \rmd r^2 - r^2 \rmd\phi^2 - \rmd z^2$ expressed in cylindrical coordinates. Therefore, we use the Cartesian-like coordinates
\begin{equation}
x^{\mu}\equiv (t,x,y,z) \equiv (t,r\cos\phi,r\sin\phi,z),
\end{equation}
as the basis for all space-time diagrams of G\"odel's universe.

The field-generating matter of the G\"odel space-time is at rest with respect to the coordinates $(r,\phi,z)$. Thus, considering an arbitrary point $P$ in space, the spatial orientation defined by this matter, which we designate as the stellar compass, coincides with the coordinate axis at $P$. However, the inertial compass at $P$ determined, for example, by a set of gyroscopes~\cite{Kaj09,Delgado02} rotates around the $z$-axis with respect to the stellar compass. The angular velocity 
\begin{equation}
\Omega_G\equiv\frac{c}{\sqrt{2}a}\equiv\frac{\sqrt{2}c}{r_G}
\label{eq:angular_velocity}
\end{equation}
of this rotation is inversely proportional to the G\"odel radius $r_G$. As a result, the G\"odel universe represents a model in which Mach's principle~\cite{Ciufolini95} is not valid. 

An important property of the G\"odel space-time is the existence of a five-dimensional group of continuous symmetries. Three of these symmetries which are of particular interest can be directly read off by the elements of the metric (\ref{eq:metric}). Indeed, a rotation in the angular variable ($\phi\rightarrow\phi+\phi_0$), a translation in time ($t\rightarrow t+t_0$) and a translation along the $z$-axis ($z\rightarrow z+z_0$) do not change the metric. The remaining two symmetries are more subtle and can be found in~\cite{Kaj04, Gra09}.

Moreover,~(\ref{eq:metric}) reveals that the G\"odel metric is the direct sum of two subspaces formed by the $(t,r,\phi)$ coordinates and the $z$-coordinate. Since the latter subspace contributes $-\rmd z^2$ to G\"odel's line element only, the G\"odel space-time can be considered as being flat in the $z$-direction. As a result, the spatial subspace of the G\"odel universe corresponds to a stack of identical layers along the $z$-axis, each representing a Lobachewski plane~\cite{Rin09}. For this reason, we often omit the $z$-coordinate in our illustrations.

\begin{figure}[htp]
\centering
\includegraphics[width=0.67\textwidth]{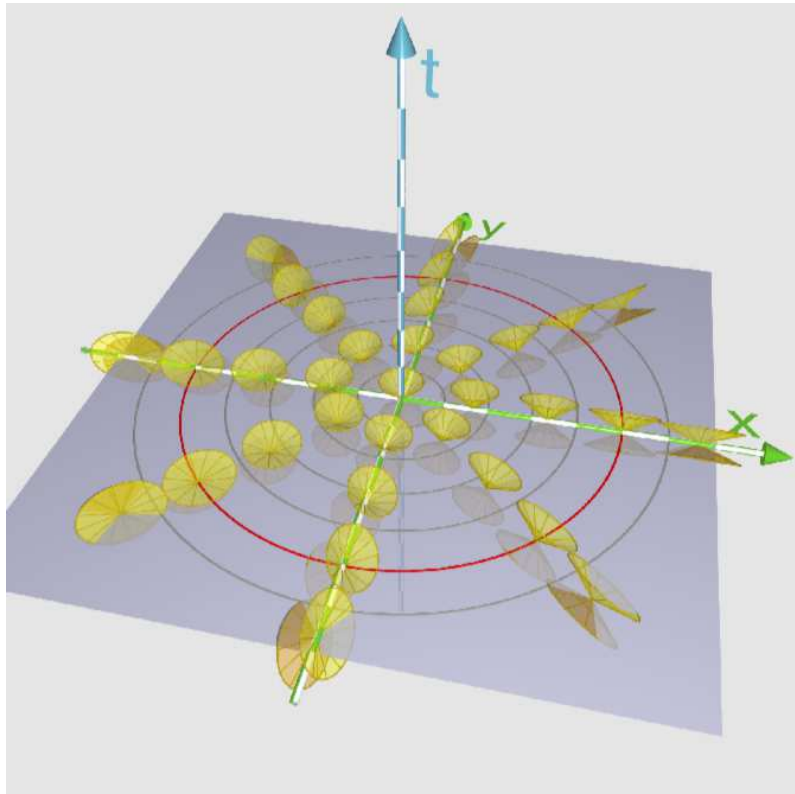}
\caption{
Light cone diagram of G\"odel's space-time (see supplementary movie 1, available from \link{stacks.iop.org/NJP/15/013063/mmedia}). The light cones are presented in the $(t,x,y)$-subspace for different $r$ and $\phi$ coordinates. Since G\"odel's space-time is stationary, the light cones do not change with $t$ and it is sufficient to consider them at $t=0$ only. This plane separates the future ($t>0$) and the past ($t<0$) defined with respect to the ideal fluid. Within the critical G\"odel radius (red circle), the future and past light cones tilt more and more with increasing separation from the origin but are still located above and below the plane, respectively. However, beyond the G\"odel radius the light cones penetrate the $t=0$ plane. Now the part of the forward light cone reaches into the domain below $t=0$.
\label{fig:lightconediagram}}
\end{figure}

In figure~\ref{fig:lightconediagram} we present the light cones for the $(x,y)$-plane with the coordinate time $t$ as vertical axis. Since the G\"odel metric (\ref{eq:metric}) is stationary and invariant under rotations, the orientation and the shape of the light cones are solely determined by the radial coordinate $r$. Light cones that are located at different points in space-time but with identical coordinates $t$ and $r$ can be merged by a simple rotation.

With increasing radial distance $r$, the light cones keel over and their opening angles widen. In the domain circumvented by the critical G\"odel radius~$r_G$, which is marked by the red circle, the future light cones are located in the positive half-space of the coordinate time $t$. However, beyond the G\"odel radius, the light cones are tilted enough that parts of their future light cones reach into the domain of negative values of coordinate time. As a consequence, an object can move in such a way that its chronological order with respect to the coordinate~$t$ is reversed. In fact, when the object moves backwards with respect to the $t$-axis, it travels into the past of an observer who pursues the world lines of the ideal fluid in the local neighbourhood of the object. While passing by, the object could observe that the watch of the observer, which displays his proper time, indeed runs backwards.

At this point it is useful to recall again the property of homogeneity of the G\"odel universe. The domain $r<r_G$ is only of physical significance as much as it is defined with respect to an observer which determines the origin of the considered coordinate system.

\subsection{Geodesics\label{subsec:geodesics}}

The worldlines of a freely falling particle as well as the curves of propagating light depend on the structure of the underlying space-time. These so-called geodesics follow from the geodesic equation
\begin{equation}
\frac{\rmd^2 x^{\mu}}{\rmd\lambda^2} + \Gamma^{\mu}_{\alpha\beta}(x^{\sigma})\frac{\rmd x^{\alpha}}{\rmd \lambda}\frac{\rmd x^{\beta}}{\rmd \lambda} = 0\,, \label{eq:geodesic}
\end{equation}
where $\lambda$ is a curve parameter that will be specified below and
\begin{equation*}
\Gamma^{\sigma}_{\lambda\mu} \equiv \frac{1}{2} g^{\nu\sigma} \left( \frac{\partial g_{\mu\nu}}{\partial x^{\lambda}} + \frac{\partial g_{\lambda\nu}}{\partial x^{\mu}} - \frac{\partial g_{\mu\lambda}}{\partial x^{\nu}} \right)
\end{equation*}
denotes the elements of the affine connection.

In order to obtain a particular geodesic as a solution of this equation, we have to provide initial conditions such as the initial position $x^{\mu}(\lambda_0)$ and the initial four-velocity $u^{\mu}(\lambda_0)$ which additionally fulfils the condition
\begin{equation}
g_{\mu\nu}(x^{\sigma}(\lambda_0))\,u^{\mu}(\lambda_0)\,u^{\nu}(\lambda_0)=\varepsilon^2,    \label{eq:condition_geodesic}
\end{equation}
where $\varepsilon\equiv c$ for time-like geodesics and $\varepsilon \equiv 0$ for light-like geodesics. The first case corresponds to worldlines of massive particles for which the curve parameter $\lambda$ represents the particle's proper time $\tau$. The latter class of geodesics, also called the null-geodesics, describes the propagation of light with $\lambda$ being an arbitrary curve parameter without a direct physical meaning.

It is not possible to solve the geodesic equation for arbitrary curved space-times. However, in the case of the G\"odel universe the symmetries mentioned in section~\ref{subsec:metric} can be exploited~\cite{Grave08,Kaj09, Gra09} to find analytic expressions for the geodesics. In the particular case of light emitted from the origin, that is $\varepsilon=0$ with initial conditions $t(0)=0$, $r(0)=0$ and $z(0)=0$, the light-like geodesics read
\numparts
\begin{eqnarray}
\fl t(\lambda)&=&-u^{(0)}(0)\,\lambda + 
\frac{2}{\Omega_G}\left( \arctan\left( \frac{\Omega_G}{\omega}\,u^{(0)}(0)\, \tan(\omega\lambda)\right) + 
\pi \left\lfloor\frac{\omega}{\pi}\lambda+\frac{1}{2}\right\rfloor \right)
\label{eq:t_lambda}\\
\fl r(\lambda) &=& 2aA\left|\sin\left(\omega\lambda\right)\right|	  
\label{eq:r_lambda}\\
\fl \phi(\lambda) &=& 
  \arctan\left( \frac{\Omega_G}{\omega}\,u^{(0)}(0)\, \tan(\omega\lambda)\right) + 
\pi\left\lfloor\frac{\omega}{\pi}\lambda + \frac12 \right\rfloor - \pi\bigg\lfloor\frac{\omega}{\pi}\lambda \bigg\rfloor + \phi(0)	
\label{eq:phi_lambda}\\
\fl z(\lambda)&=&u^{(3)}(0)\,\lambda	
\label{eq:z_lambda}
\end{eqnarray}
\endnumparts
with the constants
\begin{equation}
\omega \equiv \frac{u^{(1)}(0)}{2a}\sqrt{1+2\left(\frac{u^{(3)}(0)}{u^{(1)}(0)}\right)^2} 
\qquad \mbox{and} \qquad
A \equiv \frac{u^{(1)}(0)}{2a\omega}
\label{eq:definition_omega_A}
\end{equation}
and the floor function $\lfloor x\rfloor := \max\{n\in \mathds{Z}\,:\,n\leq x\}$.

These equations require three parameters only: the initial (azimuthal) angle $\phi\equiv\phi(0)\equiv x^2(0)$ and the two initial velocity components $u^{(1)}(0)=\frac{\rmd x^{1}(0)}{\rmd\lambda}$ and $u^{(3)}(0)=\frac{\rmd x^{3}(0)}{\rmd\lambda}$ defining the direction of the geodesic. Without loss of generality we can set $u^{(1)}(0)=1$ and use $u^{(3)}(0)=\tan\theta$ where $\theta$ denotes the angle between the null-geodesic and the $(x,y)$-plane at the origin. This simplification corresponds to a rescaling of the curve parameter according to $\lambda'=u^{(1)}(0)\,\lambda$.

\begin{figure}[ht]
\centering 
\includegraphics[width=\textwidth]{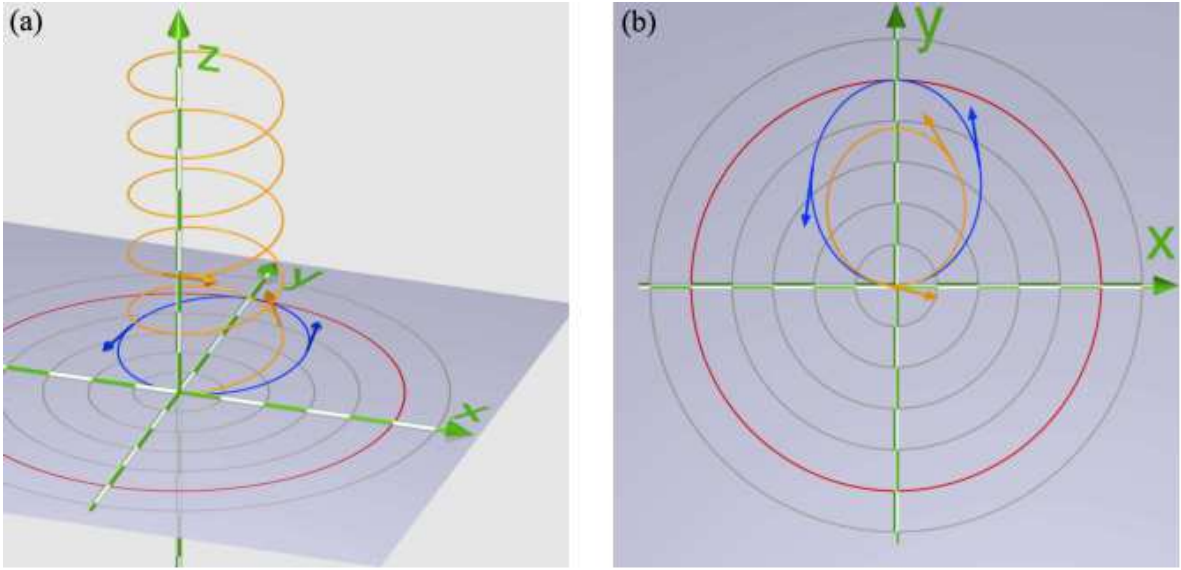}
\caption{
Null geodesics in the G\"odel universe emerging from the origin in the $(x,y,z)$-subspace (a) and in $(x,y)$-subspace (b). The red circle marks the G\"odel radius $r_G$ and defines the maximal distance a null geodesic can reach. Common to all geodesics is their helical structure aligned along the $z$-axis. Depending on the initial velocity in $z$-direction $u^{(3)}(0)$ the helix has a larger or smaller pitch at the expense of its cross section in the $(x,y)$-subspace. Geodesics for which the component $u^{(3)}(0)$ vanishes are bound to the $z=0$ plane and form closed curves of ``elliptical'' shape.
\label{fig:plot_geodesics_2d3d}}
\end{figure}

In figure \ref{fig:plot_geodesics_2d3d} we show examples of null geodesics in the $(x,y,z)$-subspace (a) and their projection into the $(x,y)$-plane (b). As shown in figure~\ref{fig:plot_geodesics_2d3d}(a), the geodesics emanate from the origin having equal initial azimuthal angle $\phi$. However, they differ in their initial elevation angles $\theta$ and thus in their initial velocity component $u^{(3)}(0)$. While the geodesic with $\theta=0$ is completely located within the $z=0$ plane due to its vanishing velocity $u^{(3)}$, the other geodesic conspicuously forms a helix along the $z$-direction. This phenomenon is a result of the Coriolis-like forces constantly deflecting the light in the $(x,y)$-subspace and the linear propagation in $z$-direction according to~(\ref{eq:z_lambda}). This motion can be alternatively interpreted as a Larmor type of motion that arises from the uniform gravitomagnetic field of G\"odel's universe as pointed out in~\cite{Costa08}. Figure \ref{fig:plot_geodesics_2d3d}(b) 
depicts the projection of the geodesics onto the $(x,y)$-subspace leading to ``elliptical'' curves.

Common to all geodesics is the fact that the curves are closed in the $(x,y)$-subspace which is a consequence of the periodicity of $r$ and $\phi$ in $\lambda$ as expressed by~(\ref{eq:r_lambda}) and (\ref{eq:phi_lambda}). Moreover, the cross section with the $(x,y)$-plane depends crucially on the initial elevation angle $\theta$. The maximal radial distance of the light ray is given by $2aA$ as predicted by~(\ref{eq:r_lambda}). The definition of the constants $\omega$ and $A$ in~(\ref{eq:definition_omega_A}) enforces the inequality $0\leq A\leq 1$ corresponding to an initial elevation angle ranging from $\pm 90^\circ$ to $0^\circ$. As a result, all light rays emanating from the origin are encompassed by a cylinder defined by the radial coordinate $r=r_G$ in the $(x,y,z)$-subspace. Only light rays with vanishing elevation angle can reach the G\"odel radius, while the ones emitted with a non-vanishing~$\theta$ possess a helix that is stretched at the expense of the ``elliptical'' cross section.

\begin{figure}[p]
\centering
\includegraphics[width=0.85\textwidth]{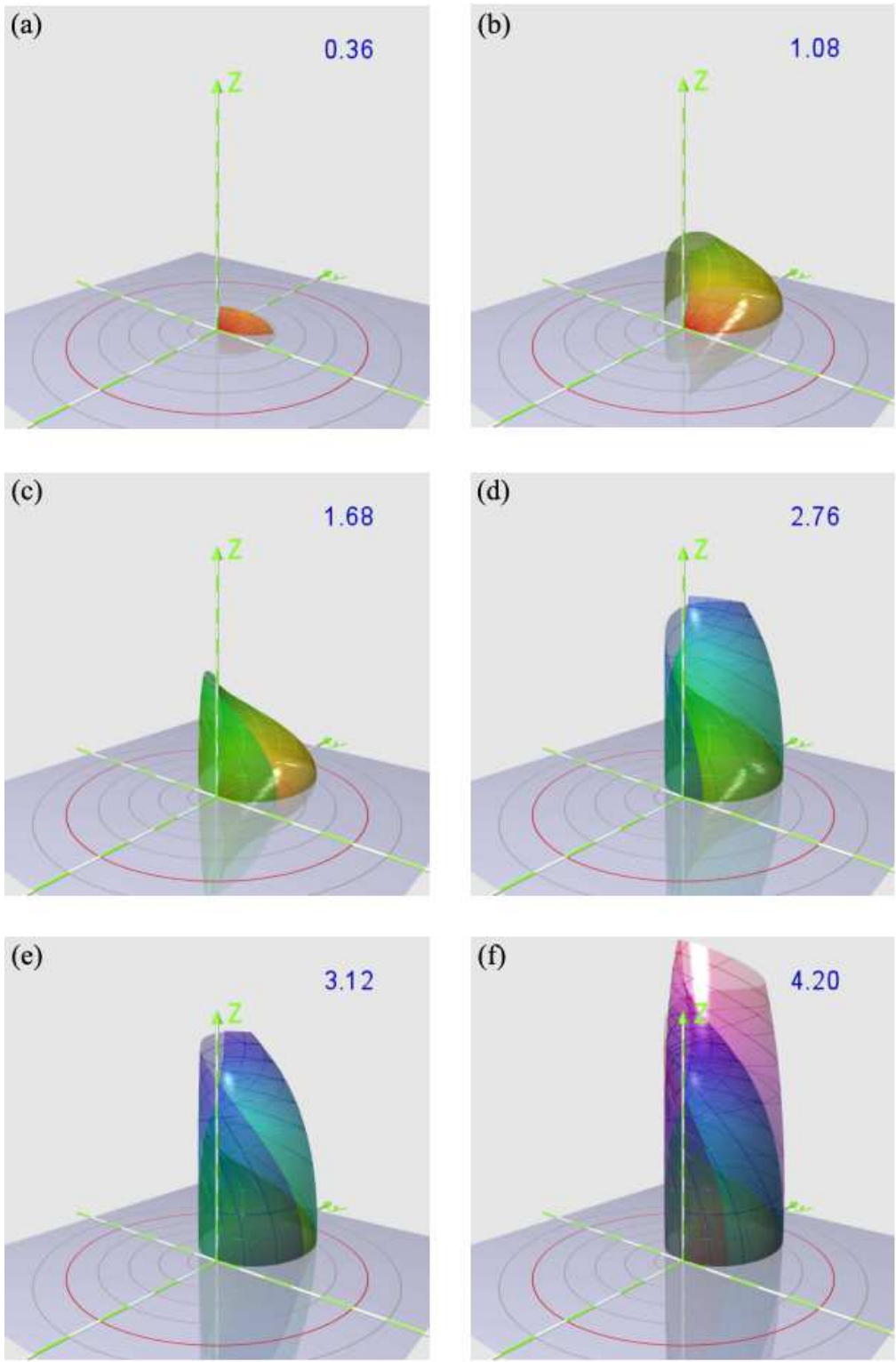}
\caption{
A bundle of null geodesics emerging from the origin and propagating in the $(x,y,z)$-subspace (see supplementary movie 2, available from \link{stacks.iop.org/NJP/15/013063/mmedia}). All geodesics have the same azimuthal angle $\phi$ and the elevation angles $\theta$ range from $-85^\circ$ to $85^\circ$. We show the geodesics for different coordinate times ranging from $t=0.36$ (a) to $t=4.20$ (f). The colour coding represents the value of the $t$-coordinate along the geodesics.
\label{fig:geodesic_fan}}
\end{figure}

We illustrate these features by looking at a bundle of null geodesics which forms a curved surface as shown in figure~\ref{fig:geodesic_fan} for the spatial $(x,y,z)$-subspace. All geodesics start from the origin at $t=0$ with the same azimuthal angle $\phi$ but with different elevation angles~$\theta$ ranging from $-85^\circ$ to $85^\circ$. Each frame presents the geodesics up to a maximal coordinate time~(\ref{eq:t_lambda}) whereby the colour indicates the value of $t$ along the geodesic bundle. The result is a symmetric structure which winds itself up and down the $z$-axis due to the helical structure of each individual light ray.
\begin{figure}[p]
\centering
\includegraphics[width=\textwidth]{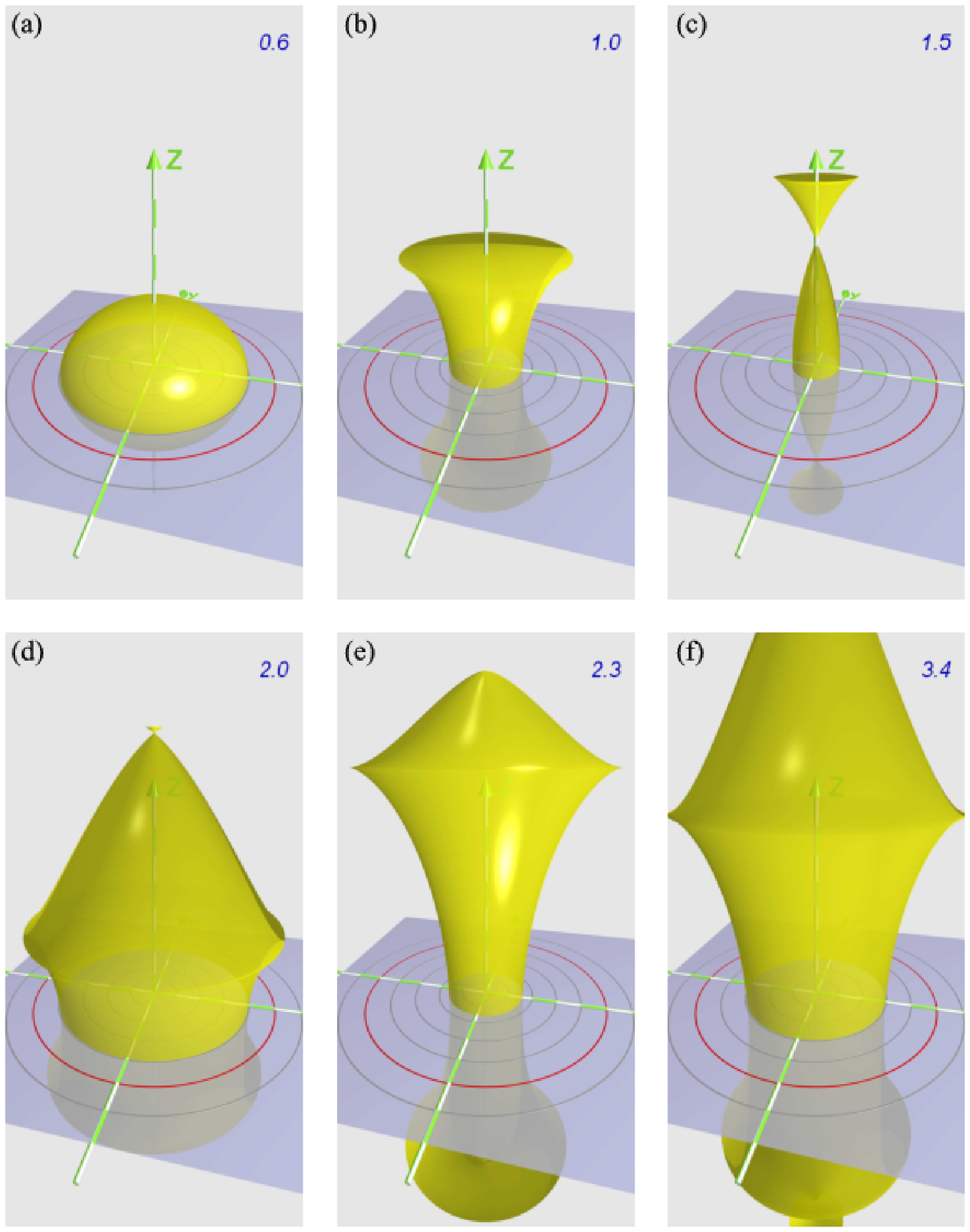}
\caption{
Wave front of a flash of light emerging from the origin and propagating through the $(x,y,z)$-subspace depicted for different coordinate times $t$ indicated in the right corner of every picture (see supplementary movie 3, available from \link{stacks.iop.org/NJP/15/013063/mmedia}). This wave front is the result of null geodesics emanating from the origin in all possible directions. The yellow transparent surface in each picture corresponds to the isosurface for a constant $t$ of all considered geodesics. 
\label{fig:lightflash}}
\end{figure}

\subsection{Ray optics\label{subsec:rayoptics}}

Next we discuss the propagation of null geodesics by means of the wave fronts they form. This topic can be summarized by geometrical optics. 

In figure~\ref{fig:lightflash} we show the wavefronts of light rays that emanate from the origin and propagate in all possible directions in the $(x,y,z)$-subspace by considering the hypersurfaces defined by the points on the geodesics with equal coordinate time $t$. These hypersurfaces correspond to light fronts of a flash dispersing through space as shown in figure~\ref{fig:lightflash} for different values of $t$. For small~$t$ as shown in figure~\ref{fig:lightflash}(a), the light front looks like an expanding sphere and is reminiscent of the situation in flat space-time in which light propagates along straight lines in all possible directions. Figure~\ref{fig:lightflash}(b) depicts the situation when the light in the $(x,y)$-plane has already bounced back once from the G\"odel radius $r_G$ marked by the red circle. Since light propagates along straight lines in the $z$-direction, the hypersurface forms a stretched axially symmetric structure restricted to a cylinder with the $r=r_G$. Moreover, due to the 
helical shape of the null geodesics the photons return to $r=0$ after a certain period that solely depends on the initial elevation angle $\theta$ at the origin. This phenomenon manifests itself in isolated punctual constrictions of the wave front located on the $z$-axis as illustrated by figure~\ref{fig:lightflash}(c). These constrictions proceed along the $z$-axis with elapsing time $t$. The figures~\ref{fig:lightflash}(d)-(f) show the propagating wave front for later times.

\subsection{Ray tracing\label{subsec:raytracing}}

For our visualizations of the G\"odel universe, we use the simple but powerful technique of ray tracing \cite{App68}. The basic idea of this concept is to trace light rays back from the observer to their origins. In standard computer graphics this is quite a simple process since only straight rays are involved. In contrast, the rays in G\"odel's universe have a helical structure due to the curvature of the underlying space-time as exemplified by the null geodesics in figure~\ref{fig:plot_geodesics_2d3d}. Since we only consider situations in which the observer is located in the origin, we can utilize the special solution~(\ref{eq:t_lambda})-(\ref{eq:z_lambda}) of the geodesic equation for the ray tracing technique.

Throughout this article we consider two different types of scenarios: (i) Stationary scenarios which are independent of the coordinate time $t$, and (ii) dynamic scenarios which vary significantly during the time the light being emitted by the objects of the scenario propagates to the observer. In contrast to the stationary case, we now have to take the propagation time of the traced light rays into account. As a result, the history of a scenario contributes decisively to its visual appearance, especially when the objects move with velocities close to the speed of light, or large distances between the objects and the observer are involved.

\subsubsection{Stationary scenarios}

\begin{figure}[ht]
\center
\includegraphics[width=0.5\textwidth]{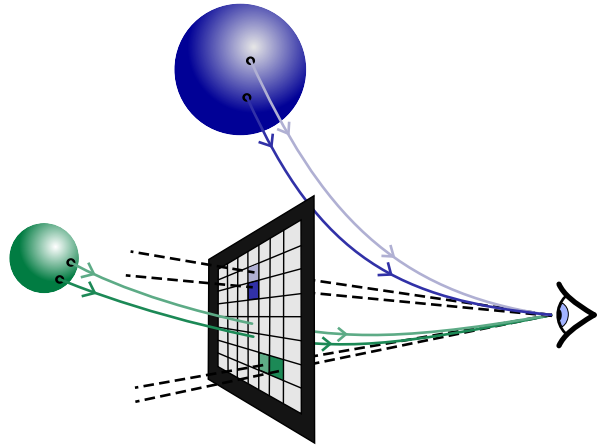}
\caption{Ray tracing of a stationary scenario in curved space-time where the light rays are bent. Since the observer has the natural notion of straight light rays, objects appear to be at positions which are different from their true locations.
\label{fig:raytracing_curved_rays}}
\end{figure}

Figure~\ref{fig:raytracing_curved_rays} illustrates the method of ray tracing for stationary scenarios in curved space-time. Light rays arriving at the observer's eye are traced back until they intersect an object being part of the visualized scenario. Since the observer is not aware of the curvature of the space-time the exact position of the object remains concealed. Indeed, the observer always perceives light to travel on straight lines as the black dashed rays suggest. This fact can be used to determine the exact position of the objects on a virtual image plane showing the visualization of the scenario. For this purpose we divide the image plane into small rectangular areas each representing a pixel of a fictitious computer screen. In order to render a visualization of a given scenario, a series of rays are constructed in a way that their imaginary straight extensions cover all pixels of the virtual screen. The colour of each pixel is thereby given by the colour of the object which is intersected by the 
corresponding ray.

\subsubsection{Dynamic scenarios\label{subsec:dynamic_scenarios}}

\begin{figure}[ht]
\center
\includegraphics[width=0.7\textwidth]{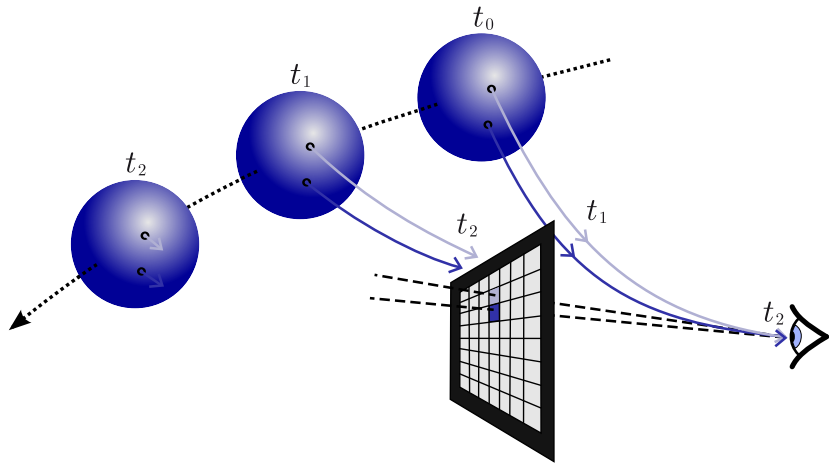}
\caption{
Timing effects of propagating light in ray tracing of a dynamical scenario. For three different times $t_0<t_1<t_2$ we exemplify the dynamics of a light emitting sphere. At the position of the sphere corresponding to $t_0$, light is emitted towards the observer. However, due to the finite speed of light it reaches him at $t_2$. While the light was propagating the sphere has already moved to a different position. An intermediate state of the sphere's motion is shown at the time $t_1$. At the time $t_2$, the light emitted by the sphere at $t_1$ has not even arrived at the screen as indicated by the curved geodesic.
\label{fig:4d_raytracing}}
\end{figure}

Figure \ref{fig:4d_raytracing} illustrates the importance of timing effects in a visualization. Here we consider a sphere which continuously sends light towards the observer while moving along a trajectory. Light which is emitted at $t_0$ from the sphere's surface reaches the observer at a later time $t_2$. Although the observer perceives the sphere as being located at its initial position, the sphere meanwhile has moved on and reached another location. As a consequence, we obtain a visualization which depends on the exact former dynamics of the scenario and, in particular, on its history. This dependence on the past gets even more mind-boggling when we think of a scenario in which light emitted at $t_2$ reaches the observer at the same time, or even before the time $t_0$ when the first light pulse was emitted.

\subsection{G\"odel horizon\label{subsec:goedelhorizon}}

The G\"odel radius $r_G$ not only determines the angular velocity of the intrinsic rotation of the space-time as given by~(\ref{eq:angular_velocity}) but also defines a boundary which light emitted at the origin can never cross. Such light rays are restricted to the cylindrical region $r\leq r_G$.

In the same way we can argue that light which stems from beyond the G\"odel radius ($r>r_G$) can never reach the observer located in the origin. This feature is a result of the fact that the G\"odel universe is homogeneous. As a consequence, the observer is unable to look beyond the G\"odel radius which acts as an optical horizon restricting the observer's view onto a cylindrical region around the $z$-axis.

Accordingly, the helical shape of the light rays depends on the value of $r_G$. The smaller the G\"odel radius, the more curved the light rays appear in order to not exceed $r_G$ in their radial coordinate. In the limit of $r_G\rightarrow\infty$ the curvature of the light rays vanishes and yields straight lines as in flat space-time.

We now address the impact of different values for $r_G$ on the visual perception of the observer. Starting from flat space ($r_G=\infty$) we stepwise decrease $r_G$ in our visualizations. As a consequence, the null geodesics change their shape from (quasi) straight lines to the typical helical structure. We demonstrate this transition by visualizing a plane in the $(x,y)$-subspace. 

In order to bring out the above mentioned optical effects most clearly, we tile the surface of the plane to be visualized with a specific pattern. The plane bears resemblance to a checkerboard where we have included additional green concentric circles around the $z$-axis in order to allude to the rotational symmetry of the G\"odel space-time. 

\begin{figure}[ht]
\centering
\includegraphics[width=0.8\textwidth]{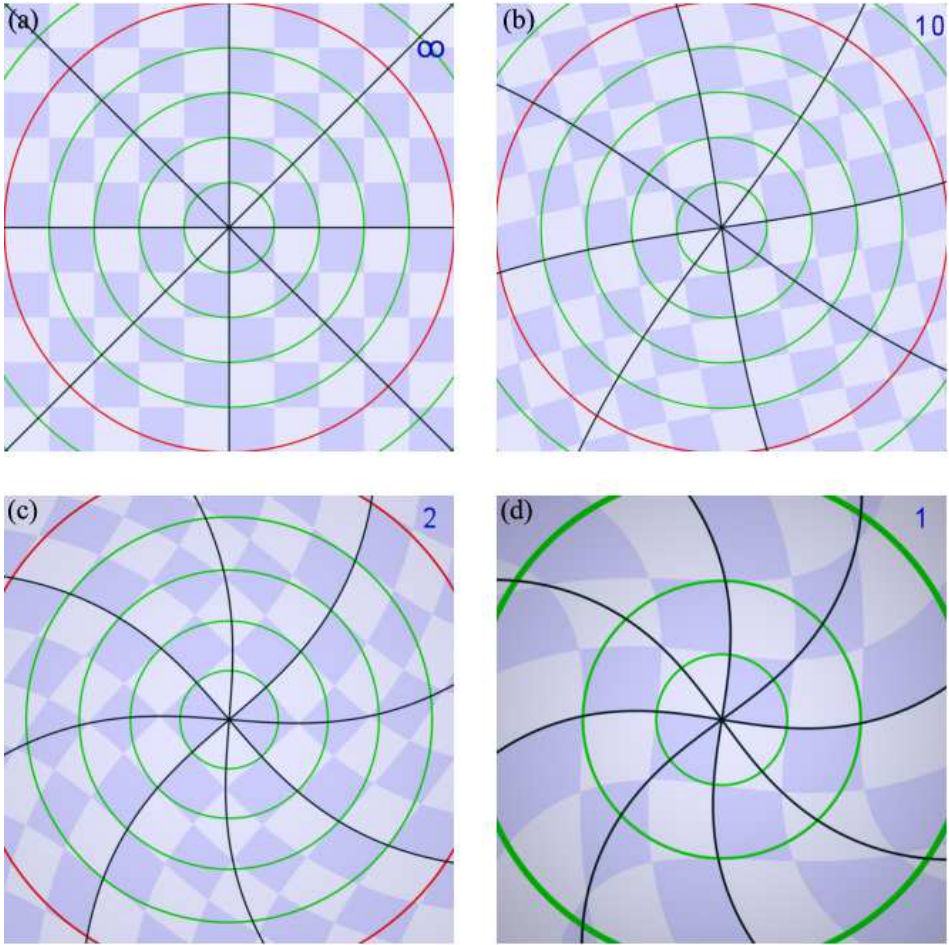}
\caption{
Intrinsic rotation of the G\"odel universe as apparent from a view onto the $(x,y)$-plane at $z=1.0$ for four different values of the G\"odel radius $r_G$ indicated on the top right corner of every frame (see supplementary movie 4, available from \link{stacks.iop.org/NJP/15/013063/mmedia}). The observer located at the origin looks straight onto the plane and has an opening angle of his view of~$90^\circ$. The checkerboard pattern of the plane is accompanied by concentric circles and black lines in order enhance the visual perception of the spacetime curvature. Figure~\ref{fig:ebene_senkrecht}(a) presents the view onto the plane in flat space-time, that is for the G\"odel radius $r_G=\infty$. With decreasing $r_G$, the plane becomes twisted as a result of the intrinsic rotation of the universe and is magnified due to the restriction of the light rays to the G\"odel radius by their helical shape.
\label{fig:ebene_senkrecht}}
\end{figure}

For our first visualization we have placed this plane at $z=1.0$. The observer looks along the positive $z$-direction and thus perpendicularly onto the plane. His visual perception is presented in figure~\ref{fig:ebene_senkrecht}. Figure~\ref{fig:ebene_senkrecht}(a) displays the undistorted plane since in this case the underlying space-time is flat. Figure~\ref{fig:ebene_senkrecht}(b) exhibits a slight twist around the centre which is enhanced in the figures~\ref{fig:ebene_senkrecht}(c) and (d). With further decreasing $r_G$, the centre of the plane is magnified which can be easily seen by counting the number of visible circles. The picture with the smallest value of the G\"odel radius depicted in figure~\ref{fig:ebene_senkrecht}(d) shows this effect vividly. Here only three green circles are visible, while the red circle of radius $1.0$ is out of sight.

\begin{figure}[ht]
\centering
\includegraphics[width=\textwidth]{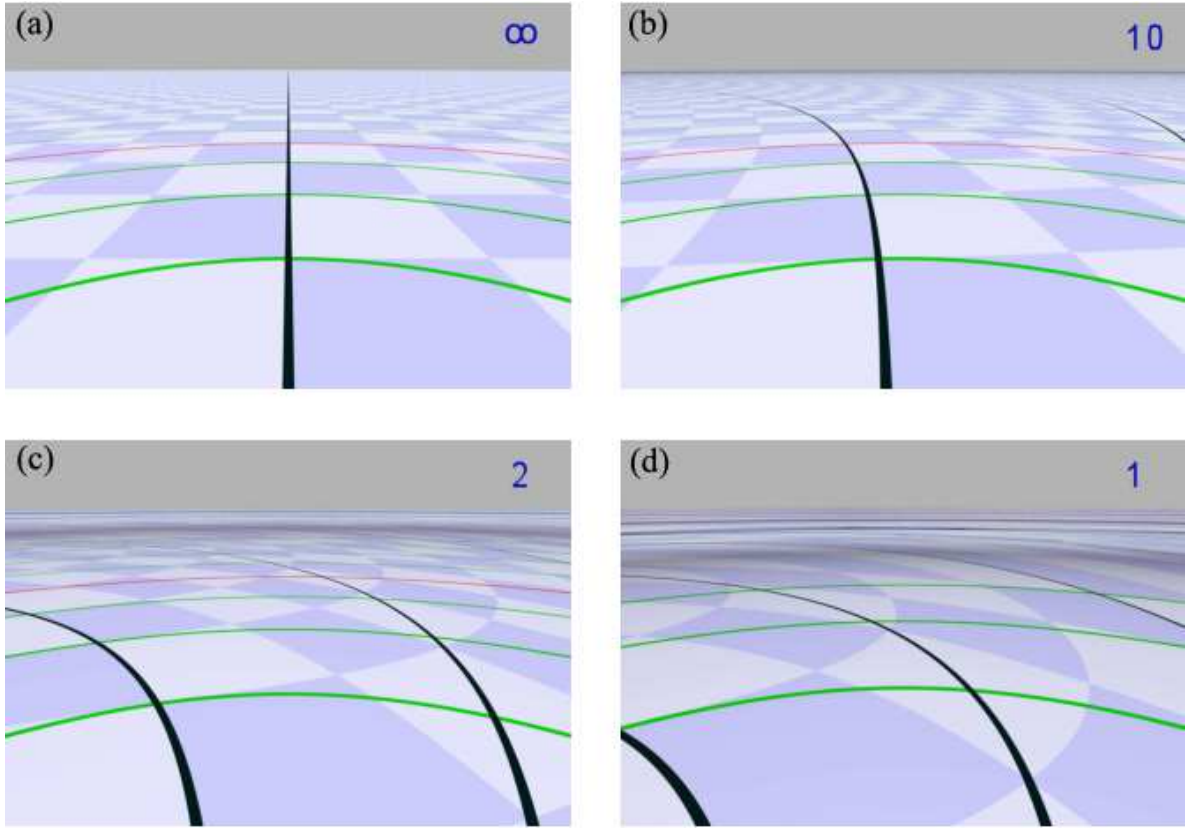}
\caption{
Influence of the intrinsic rotation of the G\"odel universe on light rays demonstrated by visualizations of the $(x,y)$-plane located at $z=-0.2$ (see supplementary movie 4, available from \link{stacks.iop.org/NJP/15/013063/mmedia}). The observer looks parallel to the plane for four different values of the G\"odel radius $r_G$ indicated in the top right corner of every frame. Figure~\ref{fig:ebene_parallel}(a) represents the situation of flat space that is $r_G=\infty$. We clearly see the tiled structure of the plane and the circles enclosing the observer. In the case of $r_G=10$ (b) the more distant parts of the plane appear to be dragged to the left due to the intrinsic rotation in G\"odel's universe. When we further decrease $r_G$ ((c) and (d)) the optical horizon at $r_G$ restricts noticeably our view. Parts of the plane which are beyond $r_G$ get out of sight and are replaced by parts of the domain inside the G\"odel horizon via higher order light rays as discussed in the next subsection.
\label{fig:ebene_parallel}}
\end{figure}

In the visualization presented in figure~\ref{fig:ebene_parallel} the $(x,y)$-plane has a position slightly below the observer at $z=-0.2$. The observer is again located at the origin and has a viewing direction parallel to the $(x,y)$-plane. Figure~\ref{fig:ebene_parallel} shows the vision for four different G\"odel radii $r_G$. Figure~\ref{fig:ebene_parallel}(a) depicts the situation for flat space-time corresponding to $r_G=\infty$. The tiled structure and the circles of the surface appear undistorted in the observer's vision. Indeed, the edges of the individual tiles as well as the depicted bar which runs towards the horizon are straight. We have clipped the top part of the picture since it is dispensable for our considerations.

Figure~\ref{fig:ebene_parallel}(b) shows the same situation for a finite value of $r_G$. Due to the incipient bending of the light rays, the structure seems to be dragged towards the left. With decreasing G\"odel radius ((c) and (d)) that effect gets more pronounced and the plane appears twisted around the observer's position. Moreover, the optical horizon in G\"odel's universe becomes noticeable. Whereas in flat space we have an unlimited view over the infinite plane, in all other cases the view is restricted to that section of the plane which is located within the G\"odel radius $r_G$. Despite the fact that distant parts of the plane with $r_G<r$ move out of sight with decreasing $r_G$, these areas are not just left blank in the visualization. Instead, they are replaced by copies of the visible region of the plane which is a result of higher order rays as discussed in the next section.

\section{Visualization of stationary scenarios\label{sec:stationary_scenarios}}

We start our series of visualizations by considering objects that are at rest. In this case, we do not need to account for timing effects resulting from the finite speed of light. We only consider scenarios consisting of spheres since their elementary mathematical description simplifies the ray tracing process considerably. The spheres we employ are very small compared to the G\"odel radius and are thus defined locally in the tangential space associated with their midpoints. In all visualizations we have chosen a G\"odel radius of $r_G=1.0$ and a value for the speed of light of $c=1.0$.

\subsection{G\"odel horizon as a mirror}

\begin{figure}[ht]
\centering
\includegraphics[width=\textwidth]{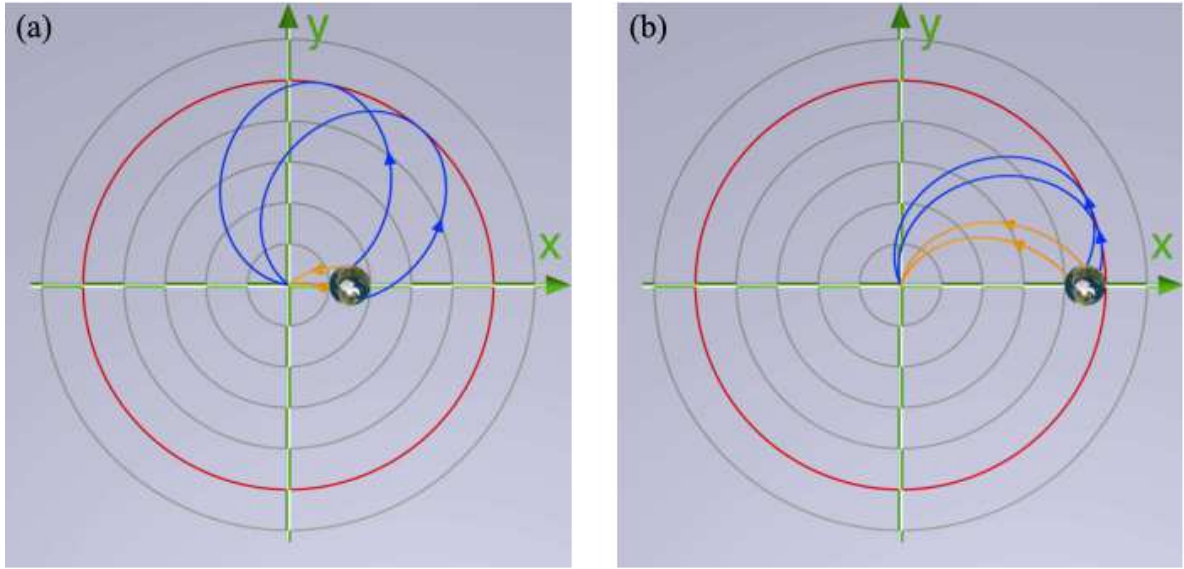}
\caption{The phenomenon of two images of an object in the G\"odel universe illustrated by the example of a terrestrial globe in the observer's view, and its dependence on the separation between observer and object. The red circle depicts the G\"odel radius. The observer is located at the origin while the globe is placed close to it (a) or near the G\"odel radius (b) in the $(x,y)$-plane. The yellow rays starting from the front of the globe propagate on the direct path while the blue rays being emitted on the back make a detour and get reflected from the G\"odel radius. They therefore travel on the indirect paths. Each ray has the typical ''elliptical'' shape of the null geodesic depicted in figure~\ref{fig:plot_geodesics_2d3d}. The two different types of paths manifest themselves in two separate images of the object displaying two different parts of the globe. Due to the ``reflection'' of the indirect rays at the G\"odel radius the resulting image is mirror-inverted.
\label{fig:strahlen_erdkugel_skizze}}
\end{figure}

In the G\"odel universe, a small object can appear twice in the view of an observer. Figure~\ref{fig:strahlen_erdkugel_skizze} brings out the physical origin of this effect and illustrates its sensitive dependence on the location of the object relative to the observer and the appearance of an optical horizon using the emission of light from a terrestrial globe.

We recognize two fundamentally different classes of light rays propagating from the globe towards the origin. On the yellow paths the light comes in a direct way from the part of the globe that faces the origin. In contrast, the blue paths make a detour and first go to the G\"odel horizon before they arrive at the origin. In this way we can observe the back side of the globe. However, due to the reflection of the light rays at the G\"odel radius, the second image of the globe is mirror-inverted. Furthermore, the different lengths of the paths make both images appear in different sizes.

\begin{figure}[p]
\centering
\includegraphics[width=0.8\textwidth]{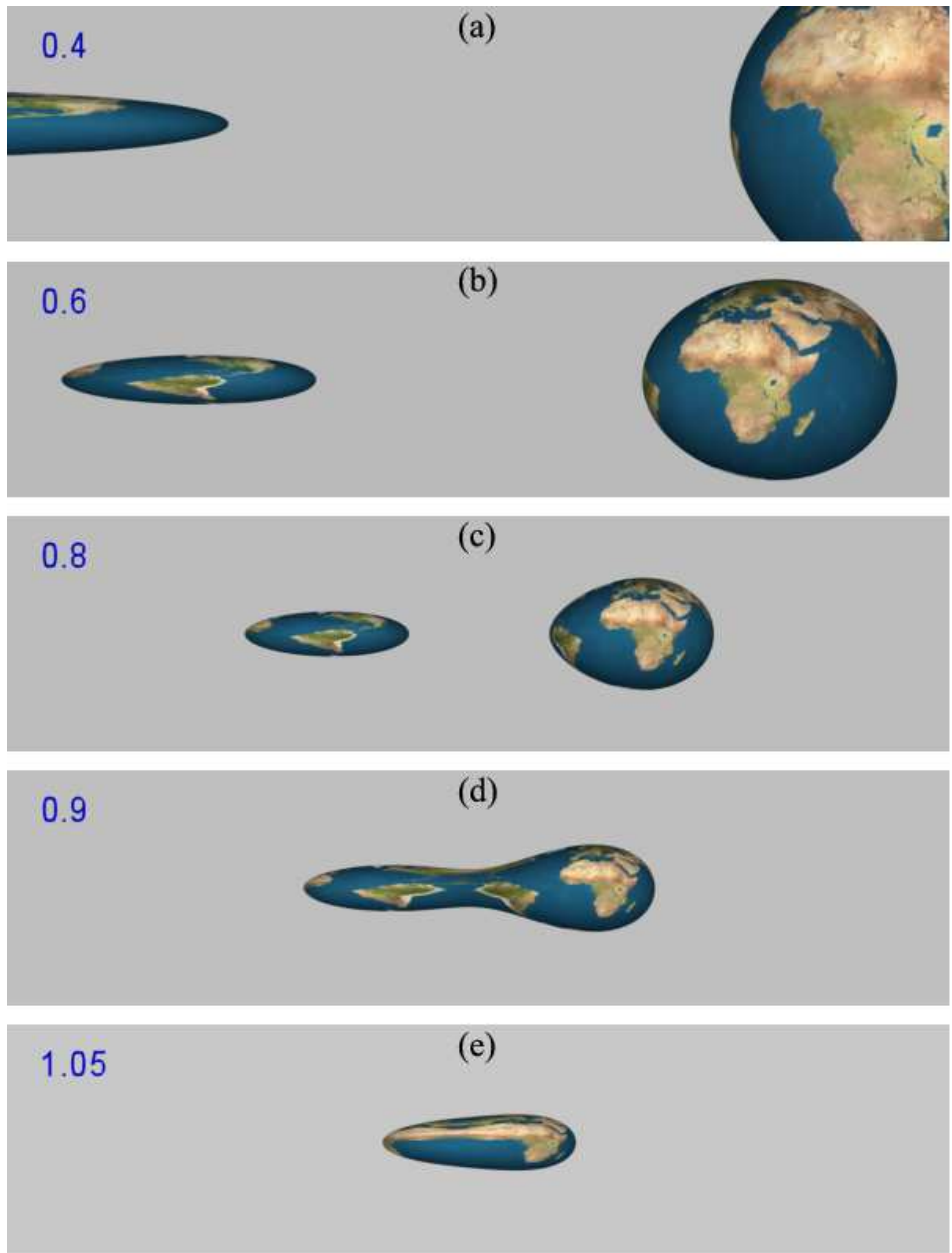}
\caption{
The G\"odel horizon as a mirror exemplified by the view from the origin on a terrestrial globe in the G\"odel universe as a function of increasing distance from the origin ((a) to (e)) scaled in units of the G\"odel radius (see supplementary movie~5, available from \link{stacks.iop.org/NJP/15/013063/mmedia}). The observer located at the origin looks in the $y$-direction with a vertical and horizontal aperture of $30^\circ$ and $120^\circ$, respectively. The globe has a radius of 0.1 and is located on the $x$-axis with a radial distance shown at the top left corner of each frame. In the figures~\ref{fig:view_erde}(a)-(c) two images of the globe are visible. The right one originates from the direct rays and the left one stems from the indirect rays and is therefore mirror-inverted and smaller in $z$-direction. Moreover, the separation of the two images is a result of the angle between direct and indirect rays in figure~\ref{fig:strahlen_erdkugel_skizze}. This angle decreases with increasing distance of the 
globe to the origin 
and eventually the two images merge as apparent in figure~\ref{fig:view_erde}(d). In figure~\ref{fig:view_erde}(e) the globe has partially crossed the G\"odel horizon and the areas of the globe cut off by the G\"odel radius are out of sight.
\label{fig:view_erde}}
\end{figure}

In figure~\ref{fig:view_erde} we now visualize the phenomenon of the two images. The globe is positioned on the $x$-axis and has a radius of 0.1. As suggested by figure~\ref{fig:strahlen_erdkugel_skizze}(b), the rays emitted by the globe arrive at the origin in opposite direction to the $y$-axis. For this reason the observer is looking in the $y$-direction with a vertical and horizontal aperture of $30^\circ$ and $120^\circ$, respectively.

Figure~\ref{fig:view_erde} shows the view of the observer for different positions of the sphere. In scenario (a) the position of the centre of the sphere is identical to the one considered in figure~\ref{fig:strahlen_erdkugel_skizze}(a). Indeed, the observer sees two images of the globe where the right one arises from the direct paths and therefore shows the front part of the sphere. In contrast, the left image is due to the indirect paths and displays the back of the globe but mirror inverted. Due to the specific helical form of the null geodesics~(\ref{eq:t_lambda})-(\ref{eq:z_lambda}) and the longer path the indirect light rays travel, the left image appears to be smaller in $z$-direction as compared to the right image of the globe.

With increasing separation of the globe from the origin (figure~\ref{fig:view_erde}(b) and (c)), the angle between the light rays corresponding to the direct and the indirect paths decreases as illustrated in the figures~\ref{fig:strahlen_erdkugel_skizze}(a) and (b). This feature manifests itself in the figures~\ref{fig:view_erde}(d) and (e) by the fact that the two images move closer together and even start to merge. In the last picture, parts of the globe are invisible since they are already located beyond the G\"odel radius.

\subsection{Multiple images}

\begin{figure}[h]
\centering
\includegraphics[width=0.71\textwidth]{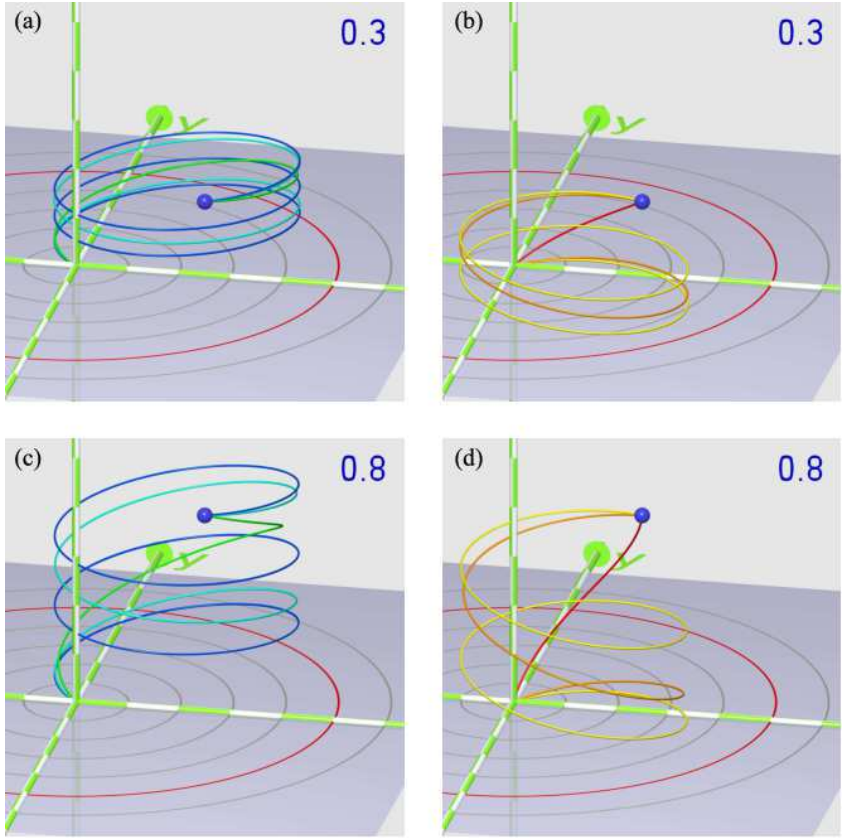}
\caption{
Multiple images created by multiple revolutions of light rays around the $z$-axis. They emanate from two different points (blue spheres) in the $(x,y,z)$-subspace and propagate along the direct path ((b) and (d)) and the indirect path ((a) and (c)) towards the origin. The two points have the spatial coordinates $r=0.5$, $\phi=0^\circ$ with $z=0.3$ ((a) and (b)) and $z=0.8$ ((c) and (d)). Direct rays propagate on the direct path towards the centre of the coordinate system ($r=0$), whereas the indirect rays make first a detour to the G\"odel horizon, which is indicated by the red circle in the $(x,y)$-plane, before they arrive in the centre. Furthermore, we call the number of completed revolutions of a ray, before it passes the origin, the {\it order} of the light ray. The red and the green rays in the pictures are rays of zeroth order since they reach the origin in their first revolution. For the sake of clarity we only show rays up to second order in all pictures. However, the order of light rays is in 
general not limited.
\label{fig:strahlengang_3d}}
\end{figure}

Next we consider another surprising optical phenomenon typical for the G\"odel universe. Spheres above or below the observer's $(x,y)$-plane repeat themselves essentially infinitely many times in the observer's vision. This intriguing effect is a result of the helical structure of the null geodesics shown in figure~\ref{fig:strahlengang_3d}. 

The little blue sphere in each frame represents a particular point on the surface of the globe to be viewed by the observer. The pictures in each row of the figure show the same position of the sphere. In complete accordance with figure~\ref{fig:strahlen_erdkugel_skizze}, we distinguish direct and indirect light rays depicted by the red, orange and yellow {\it direct} rays ((b) and (d)) and by the green, blue-green and blue {\it indirect} rays ((a) and (c)). Here we have used on purpose similar colours as in figure~\ref{fig:strahlen_erdkugel_skizze}. Since all null geodesics are helices, the light can revolve an arbitrary number of times before it finally reaches the origin. We designate rays which have completed $n$ revolutions before they arrive at the origin as rays of {\it $n$-th order}. Hence, in this classification the red and green rays of both columns in figure~\ref{fig:strahlengang_3d} which hit the observer in their first revolution are zeroth order rays.

Every light ray in figure~\ref{fig:strahlengang_3d} starts with an individual initial velocity $u^{(3)}(0)$. The pictures bring out the following feature: the smaller $u^{(3)}(0)$ and thus the smaller the elevation angle $\theta$ is, the more revolutions the light has to conduct. Moreover, the main direction from which the light ray hits the observer, that is the azimuthal angle $\varphi$ is determined by the type, direct or indirect, of the light ray. These subtle geometrical relations give rise to the peculiar distribution of the intersection points of the rays on the image plane and manifest themselves in a multiple appearance of the object in the observer's view. 

\begin{figure}[ht]
\centering
\includegraphics[width=\textwidth]{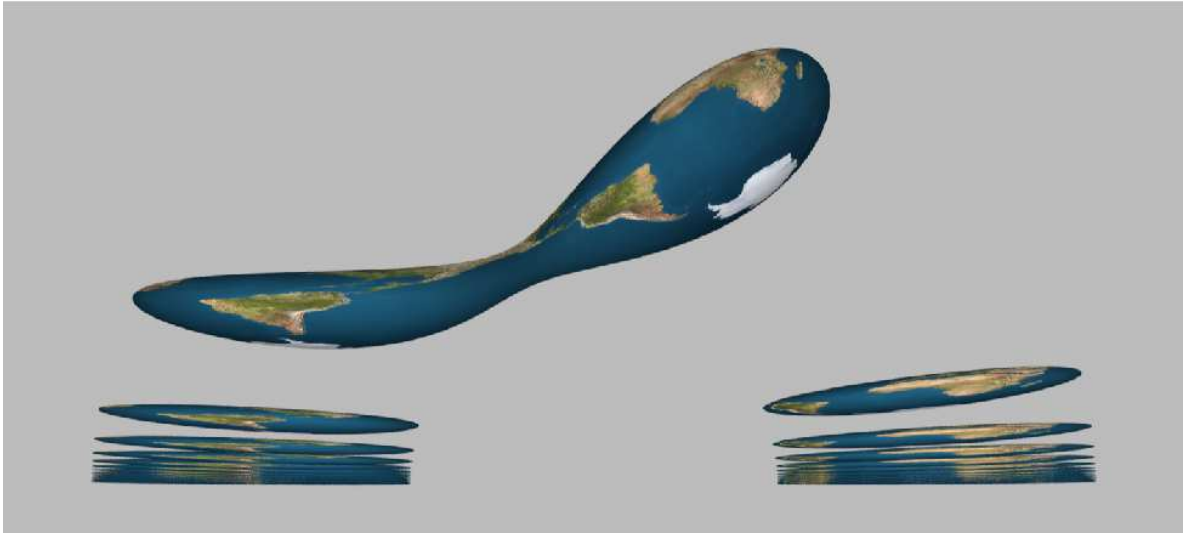}
\caption{
Multiple images in the G\"odel universe as illustrated by a visualization of a terrestrial globe that in contrast to the one in figure \ref{fig:view_erde} is slightly displaced in the positive $z$-direction (see supplementary movie 6, available from \link{stacks.iop.org/NJP/15/013063/mmedia}). Its coordinates are $r=0.74$, $\phi=0^\circ$ and $z=0.8$, the horizontal and vertical apertures are $100^\circ$ and $90^\circ$, respectively. The observer is looking in $y$-direction. The vertical displacement of the globe results in multiple images of itself. A special feature in this scenario is the merging of the images arising from the direct and indirect rays of zeroth order (top) due to the extension of the globe. We have clipped the bottom part of the visualization since no images appear below the observer's horizon.
\label{fig:view_erde_vert_displaced}}
\end{figure}

Figure~\ref{fig:view_erde_vert_displaced} displays the situation for a terrestrial globe. The two images on the top in the picture arise from the direct and indirect zeroth order rays. %
They appear spatially close to each other and even merge due to the relative large separation $r=0.74$ of the globe from the observer. In contrast the images resulting from light rays of higher order generate the stack of additional images that agglomerate near the horizon of the $(x,y)$-plane.%

\subsection{Effect of anisotropy}

In this section we rotate the observer's field view by $90^\circ$ so that his viewing direction is now aligned with the positive $z$-axis. Since the G\"odel universe is non-isotropic, we expect crucial differences in the visual appearance of objects compared to the previous visualizations where the observer looked parallel to the $(x,y)$-plane.

\subsubsection{Single sphere on the $z$-axis}
\enlargethispage*{2ex}

\begin{figure}[ht]
\centering
\includegraphics[width=\textwidth]{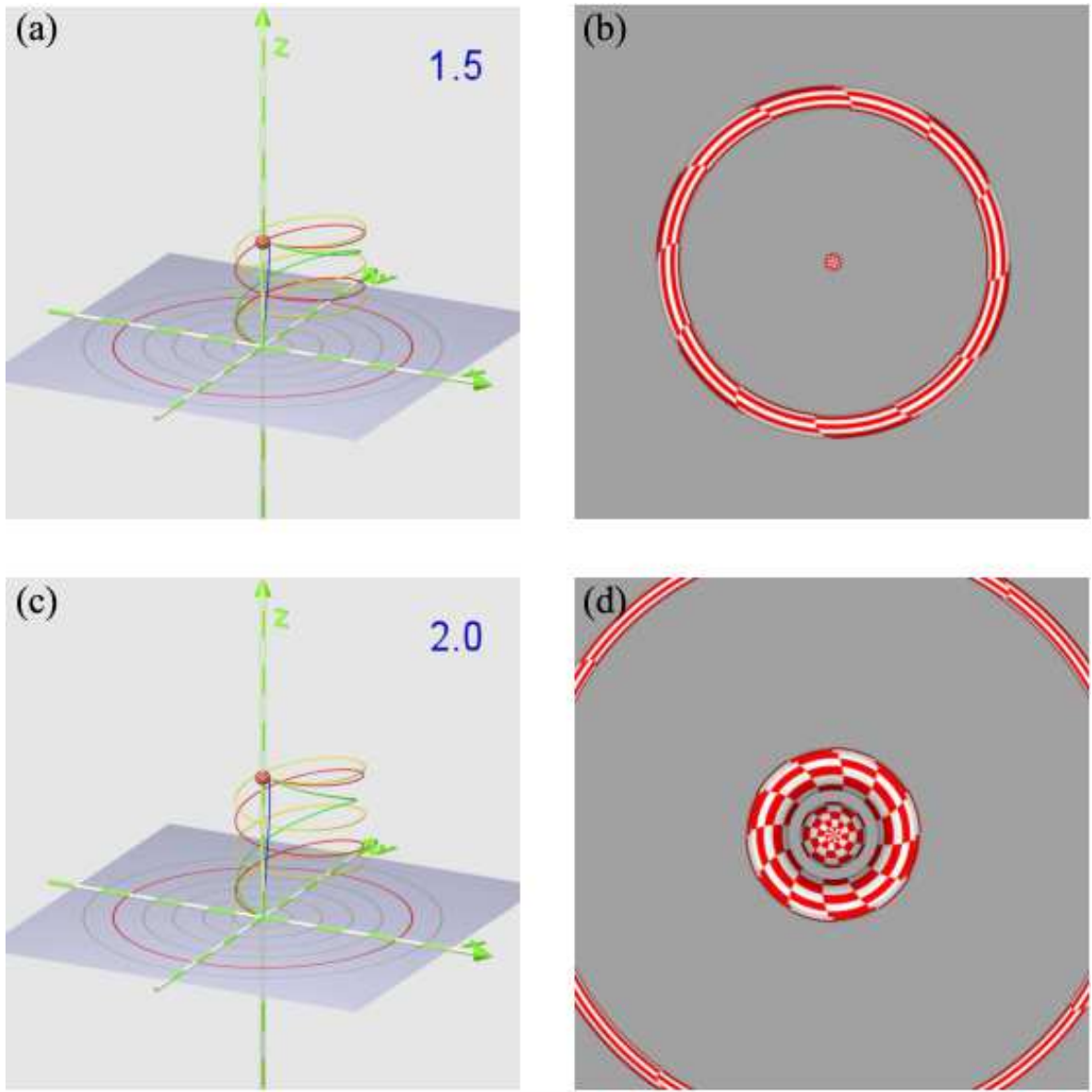}
\caption{
Effect of the anisotropy of the G\"odel universe brought to light by visualizations ((b) and (d)) of a checkered sphere located at two different positions on the $z$-axis ((a) and (c)) (see supplementary movie 7, available from \link{stacks.iop.org/NJP/15/013063/mmedia}). The positions are indicated by the value of the $z$-coordinate on the right top corner of each frame. The observer is looking along the $z$-direction. In the illustrations (a) and (c), the light rays originate from the same point on the sphere's equator. The blue ray depicts the zeroth order direct ray and the green, magenta and yellow rays represent a first, second and third order direct ray, respectively. The primary images of the sphere are located in the centre of the picture. Images which arise from higher order rays have a circular shape of different diameter.
\label{fig:kugel_z_achse}}
\end{figure}

In figure~\ref{fig:kugel_z_achse} we present the visualization of a single sphere with a checkered surface located at two different positions on the $z$-axis shown in the left column ((a) and (c)). %
Here, we also exemplify a few low order light rays emitted from the sphere that reach the observer. The corresponding visualizations are presented in the right column ((b) and (d)).

In each picture the image of the sphere is clearly seen in the centre. This image is a result of direct rays of zeroth order depicted by the blue ray on the left. They are hardly influenced by the space-time structure since they mainly propagate along the $z$-direction in which the G\"odel universe does not display curvature. However, as already exemplified by the figures~\ref{fig:strahlengang_3d} and \ref{fig:view_erde_vert_displaced}, there also exist paths of higher order. Due to the rotational symmetry of the G\"odel space-time around the $z$-axis, the additional images of the sphere appear as rings around the centre. The diameter of each ring is determined by the incident angle of the light ray at the observer which is a sensitive function of the distance between the sphere and the origin.

\subsubsection{Stack of spheres}
\begin{figure}[ht]
\centering
\includegraphics[width=0.9\textwidth]{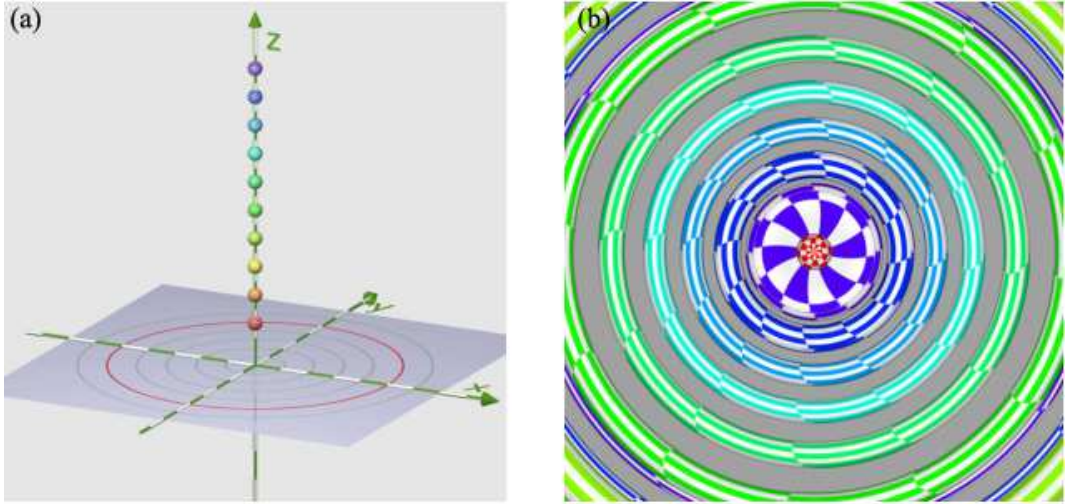}
\caption{
Rotational symmetry of G\"odel universe illustrated by the visualization (b) of a series of checkered spheres located on the $z$-axis (a). The observer in this scenario looks along the positive $z$-direction. The first sphere (red) is at $z=0.3$ and the following ones have a separation of 0.2. 
\label{fig:kugelstapel_z}
}
\end{figure}

\enlargethispage*{4ex}
Finally, we visualize in figure~\ref{fig:kugelstapel_z} a series of checkered spheres. Figure~\ref{fig:kugelstapel_z}(a) shows the arrangement of the spheres all of which have a different colour. Figure~\ref{fig:kugelstapel_z}(b) presents the visualization of this particular scenario. One particularly interesting feature is that unlike the green and blue spheres, the yellow and orange ones are not visible in the visualization. This is due to the fact that the zeroth order light rays that connect the yellow and orange spheres with the observer hit the red sphere before they reach the origin.\enlargethispage*{4ex} Moreover, the corresponding first and higher order light rays arrive at the observer at an incident angle that lies outside the opening angle of his view.

\begin{figure}[ht]
\centering
\includegraphics[width=0.9\textwidth]{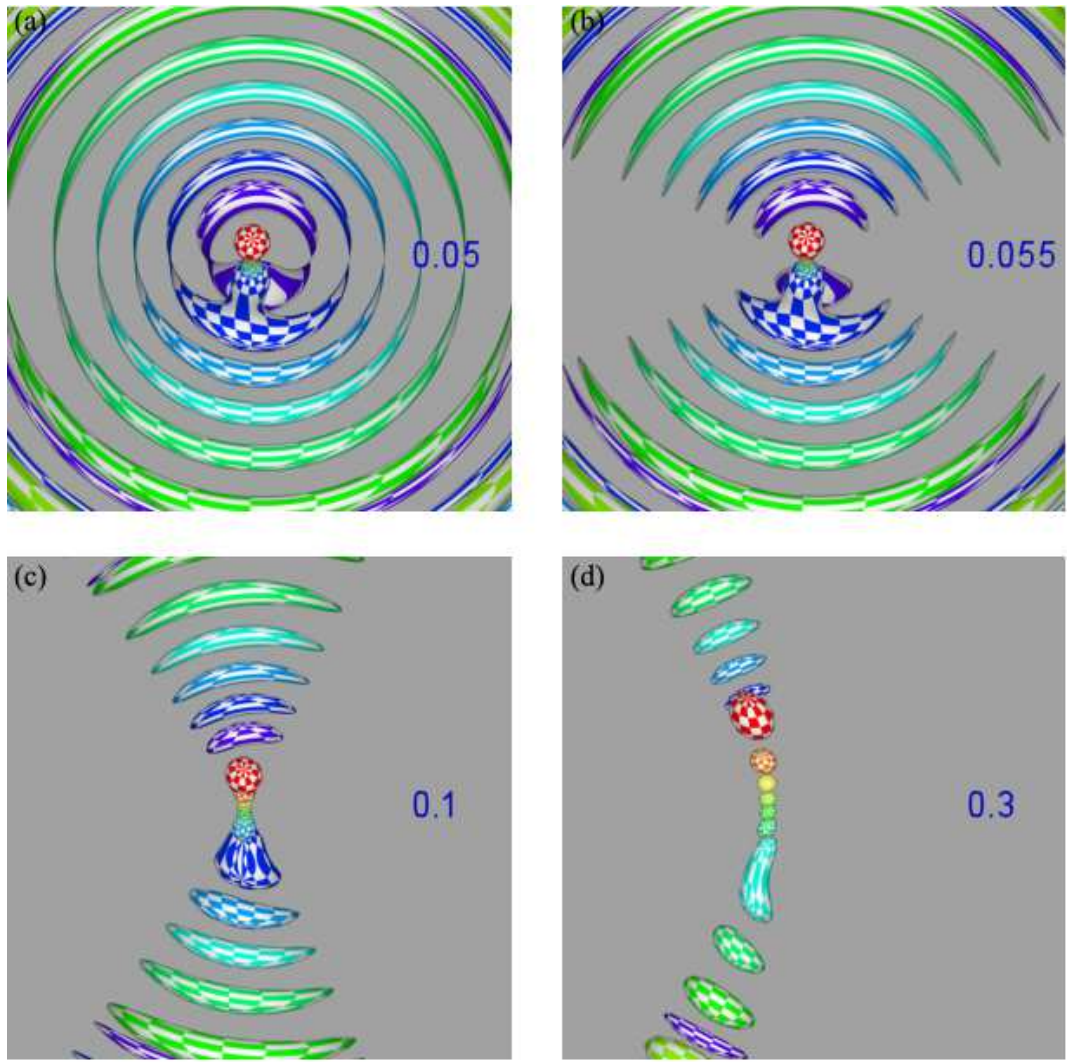}
\caption{
Break-up of rotational symmetry in the view along the $z$-direction on a stack of spheres which are slightly displaced into the $x$-direction (see supplementary movie 8, available from \link{stacks.iop.org/NJP/15/013063/mmedia}). The amount of displacement is denoted in the right middle of each frame. With increasing radial distance, the circular image of each sphere breaks up into two separate parts which are formed by the direct and the indirect rays in complete analogy to figure \ref{fig:view_erde}.
\label{fig:kugelstapel_z_achse_verschoben}}
\end{figure}

In figure~\ref{fig:kugelstapel_z_achse_verschoben} we again visualize the stack of spheres of figure~\ref{fig:kugelstapel_z}, but now with a slight displacement into the $x$-direction. Therefore, each sphere has a non-vanishing radial coordinate $r$. With increasing radial distance, the circular structure of each sphere breaks up into two separate images reminiscent of the two images of the terrestrial globe in figure~\ref{fig:view_erde}. Indeed, the two images have their origin in the direct and indirect light rays. %

\section{Visualization of freely falling spheres\label{sec:dynamic_scenarios}}

In order to demonstrate the timing effects which arise from the fact that light propagates from the object to the observer with a finite speed as outlined in section \ref{subsec:dynamic_scenarios}, we now visualize a freely falling sphere. The trajectory of such a sphere is described by a time-like geodesic~\cite{Gra09}, which has a helical shape similar to the null geodesics shown in figure~\ref{fig:plot_geodesics_2d3d}. The cross section of this helix with the $(x,y)$-subspace is determined by the initial velocity component in this subspace. 

\begin{figure}[p]
\begin{center}
\includegraphics[width=\textwidth]{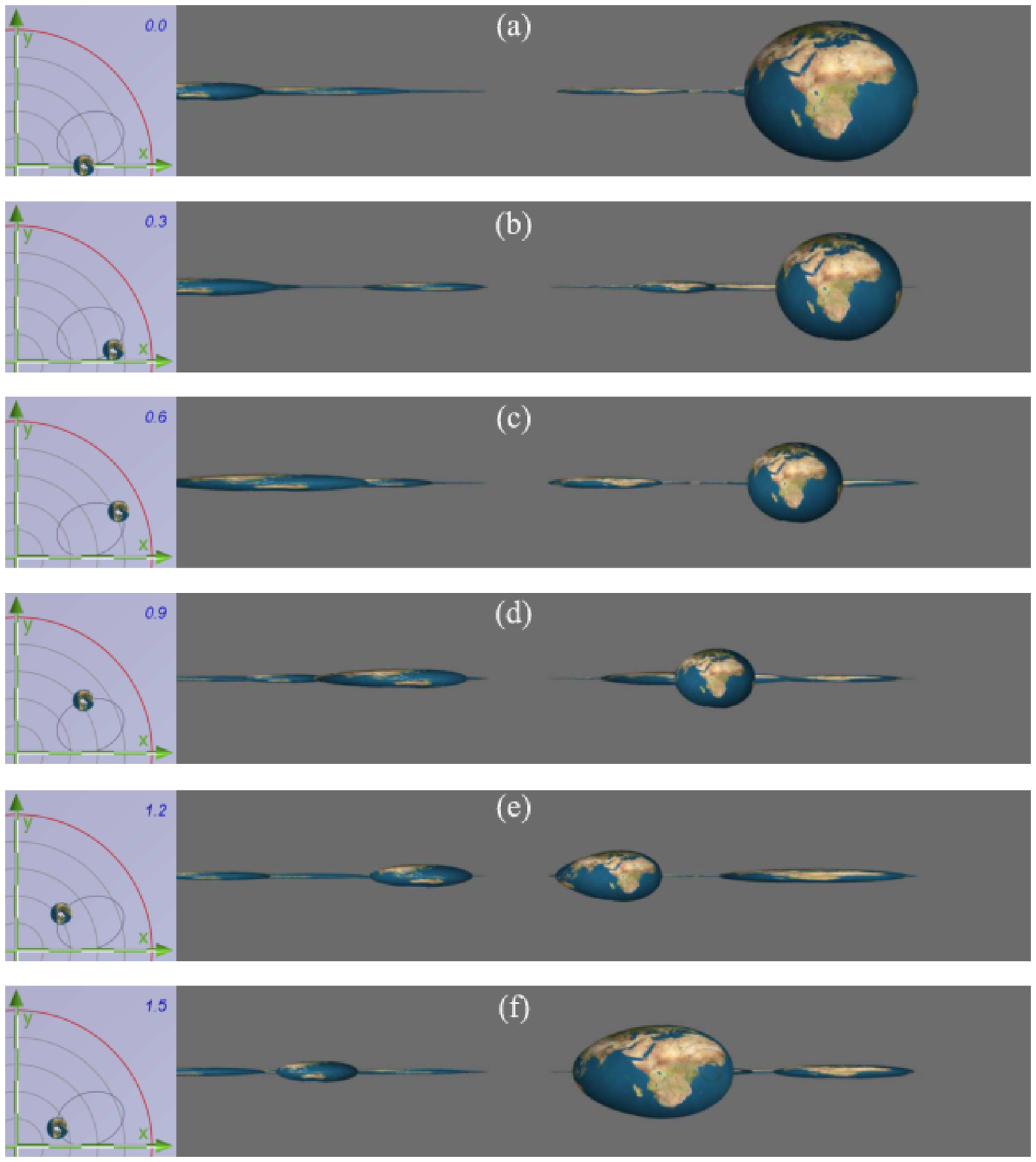}%
\end{center}
\caption{
Visualization of a moving terrestrial globe (see supplementary movie 9, available from \link{stacks.iop.org/NJP/15/013063/mmedia}). The observer is located at the origin and looks along the $y$-direction parallel to the $(x,y)$-plane with horizontal and vertical apertures of $200^\circ$ and $50^\circ$, respectively. The globe moves on a time-like geodesic within the $(x,y)$-plane which has the shape of an ''ellipse'' (left pictures). For $t=0$ the globe is released at the coordinates $x=0.5$, $y=0$ and $z=0$ with the initial velocity $u^{(1)}(0)=0.8$, $u^{(2)}(0)=0$ and $u^{(3)}(0)=0$. As a result, the globe moves first away from the observer until it reaches a turning point and travels back to the position where it started from. Each picture on the right shows a snapshot of the corresponding scenario as seen by the observer at the coordinate time~$t$ which is identical to his proper time.
\label{fig:dyn_erde_geo1}}
\end{figure}

In the scenario shown in figure \ref{fig:dyn_erde_geo1}, we consider a sphere with a vanishing velocity component in $z$-direction. Since the sphere is released at $z=0$ its trajectory describes a closed curve in the observer's $(x,y)$-plane. The left pictures in figure~\ref{fig:dyn_erde_geo1} show the terrestrial globe moving on such a geodesic. They present the $(x,y)$-subspace with the position of the globe on this geodesic (grey ''ellipses'') at the particular coordinate time $t$. The part of the red circle marks the G\"odel horizon. The pictures on the right show the corresponding visualizations of the globe.

\begin{figure}[p]
\centering
\includegraphics[width=\textwidth]{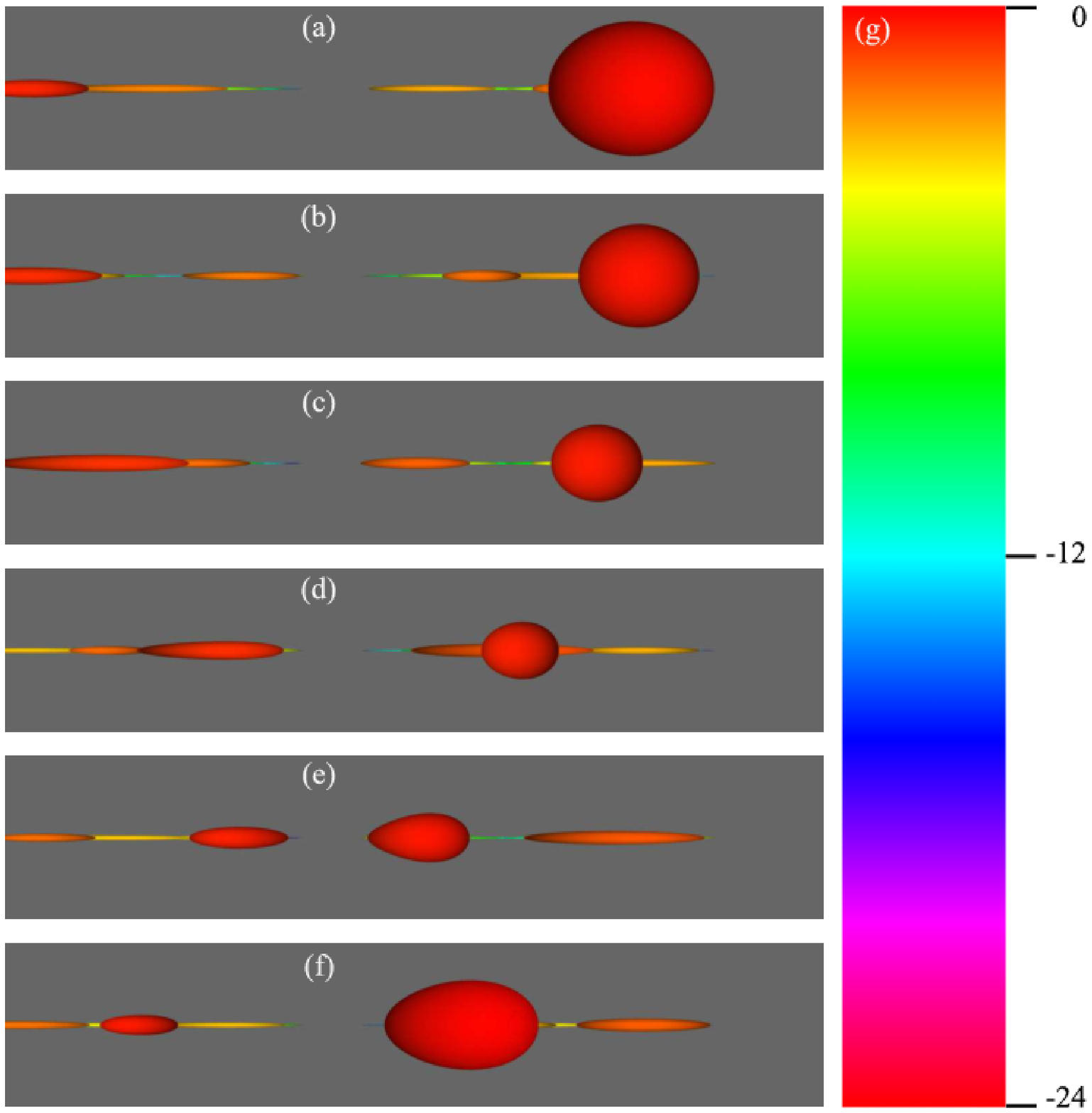}
\caption{
Visualization of temporal retardation effects for the moving globe scenario of figure~\ref{fig:dyn_erde_geo1} (see supplementary movie 9, available from \link{stacks.iop.org/NJP/15/013063/mmedia}). Each frame (a)-(f) shows the same situation as the corresponding one in figure~\ref{fig:dyn_erde_geo1}. The light rays are emitted by the globe at its proper time $\tau_\grm$ and reach the observer at his proper time $\tau=t$. For the sake of simplicity, we associate $\tau_\grm=0$ with the observer's time $\tau=0$ of figure~\ref{fig:dyn_erde_geo1}(a). In order to illustrate the temporal retardation effects, we have coated the globe's surface with a single colour. The colour scale (g) assigns to each colour a value for the proper time difference~$\tau_\grm-\tau$. The colour scale suggests that images which display a shrunken sphere in $z$-direction (corresponding to higher order light rays) appear green or even blue.
\label{fig:dyn_erde_geo1_eigentime}
}
\end{figure}

In contrast to figure~\ref{fig:view_erde} where the globe is at rest, we now see a variety of images distributed along the observer's horizon. The two images of the globe in figure~\ref{fig:view_erde} arise from the direct and indirect rays of zeroth order. Higher order rays do not contribute to the visualization since they hit the globe due to their helical structure before they arrive at the observer. However, in case of the moving globe depicted in figure~\ref{fig:dyn_erde_geo1} the globe changes its position while the light rays propagate along their individual helices. Hence, light paths which were once obstructed are cleared and thus light rays of higher order can reach the observer. As a result, with each type of light ray another image of the globe appears in the visualization.

While propagating along the various direct and indirect light rays, the light needs different ''amounts of time'' to reach the observer after being emitted at the sphere's surface, as illustrated in figure \ref{fig:4d_raytracing}. Therefore, each image shows the globe at a different proper time $\tau_\grm$ along its time-like geodesic. Since the globe's surface remains unchanged as a function of $\tau_\grm$, this phenomenon is not discernible in figure~\ref{fig:dyn_erde_geo1}. In order to make this temporal retardation effect visible, we adapt our scenario by coating the terrestrial globe with a single colour which depends on the sphere's proper time~$\tau_\grm$. 

The figures~\ref{fig:dyn_erde_geo1_eigentime}(a)-(f) show the corresponding visualization with a colour scale~(g) that depends on the difference $\tau_\grm-\tau$ between the proper time of the globe~$\tau_\grm$ at which the light is emitted and the proper time of the observer~$\tau$ at which the light is detected. The red colour marks the light emitted most recently and mainly corresponds to the two images familiar from the static case demonstrated in figure~\ref{fig:view_erde}. In contrast, the green and blue coloured images show the former sphere. Their extension in $z$-direction is significantly smaller since they stem from higher order light rays.

\section{Visualization of time travelling\label{sec:time_traveling}}

As mentioned in the introduction, the G\"odel universe allows for CTCs and thus provides the possibility to travel back in time. However, such a journey cannot be realized by free fall only. Indeed, a proper acceleration \cite{Gra09} has to be applied to the object and thus, a CTC does not represent a geodesic in G\"odel's universe.

\begin{figure}[h]
\centering
\includegraphics[width=0.7\textwidth]{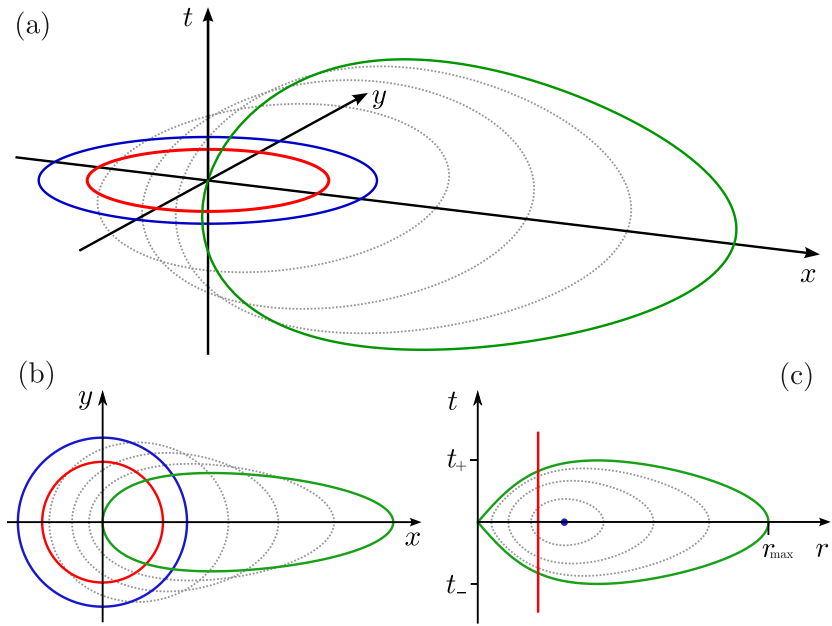}
\caption{
Transformation of a concentric CTC (blue solid line) to a CTC (green solid line) passing the origin illustrated in the $(t,x,y)$-subspace (a), in the $(x,y)$-subspace (b), and in the $(t,r)$-subspace (c) (see supplementary movie 10, available from \link{stacks.iop.org/NJP/15/013063/mmedia}). Some intermediate CTCs which result from the transformation are represented by dotted curves. In figure~\ref{fig:ctc}(a) and~(b) the red circles denote the G\"odel radius. In figure~\ref{fig:ctc}(c) it degenerates to a vertical red line. The interval $[t_-;t_+]$ corresponds to the time interval covered by the transformed CTC, and the blue dot indicates the radius of the initial concentric CTC.
\label{fig:ctc}}
\end{figure}

The most elementary version of a CTC is given by a circle around the $z$-axis with radius $r_0$ and constant coordinate time~$t$. In order to be a CTC, the radius $r_0$ has to be larger than the G\"odel radius $r_G$. An example of such a worldline is given by the blue solid circle depicted in figure~\ref{fig:ctc}(a) which has a radius of $1.4\,r_G$. The red circle marks again the G\"odel radius~$r_G$.

Since this CTC does not cross the observer's optical horizon, an object moving on such a CTC is invisible to the observer. For our purpose, we want a CTC which passes through the origin such that the object can start its journey at the observer's position. Since the G\"odel universe is homogeneous, such CTCs exist. We obtain them by making use of an appropriate isometric transformation~\cite{Gra09} which maps the entire concentric CTC onto a CTC that passes through the origin as shown by the green solid line in figure~\ref{fig:ctc}(a). The result is a CTC stretched along the $x$-axis. The intermediate stages of the transformation are indicated by the dotted curves and are themselves CTCs. Note that for all curves the $z$-coordinate is kept constant. For the visualization we choose $z=0$ for the CTC such that it lies in the observer's $(x,y)$-plane. Figure~\ref{fig:ctc}(b) represents the projection of these curves in the $(x,y)$-subspace.

Figure~\ref{fig:ctc}(c) displays these curves in $r$- and $t$-coordinates and shows that on such a CTC, a time traveller starting at $t=0$ first proceeds into the ''future'' to the coordinate time $t_{+}$ before he travels into the ''past'' all the way back to the time $t_{-}$. In this way he crosses twice the G\"odel radius which is indicated by the red vertical line.

It is the green CTC in figure~\ref{fig:ctc} which we use to visualize time travel. Since the whole CTC is restricted to a finite interval in $t$, the observer can see the object travelling along the CTC only for a limited time. We now face a puzzling question: The CTC is a {\it closed} worldline and therefore one could image that the object travels periodically along the CTC while its proper time elapses as usual. Such a situation would result in a serious problem from the observer's point of view. Indeed, the object passes a given point in space-time such as the origin, infinitely many times when it travels periodically on the CTC. Each time the object returns to this space-time point, its proper time $\tau_\orm$ has increased. For the observer, however, it is always the same event. The object would exist in infinitely many versions at the same position and at the same coordinate time~$t$, even though each version would differ in its proper time $\tau_\orm$.

\begin{figure}[htp]
\begin{center}
\includegraphics[width=0.68\textwidth]{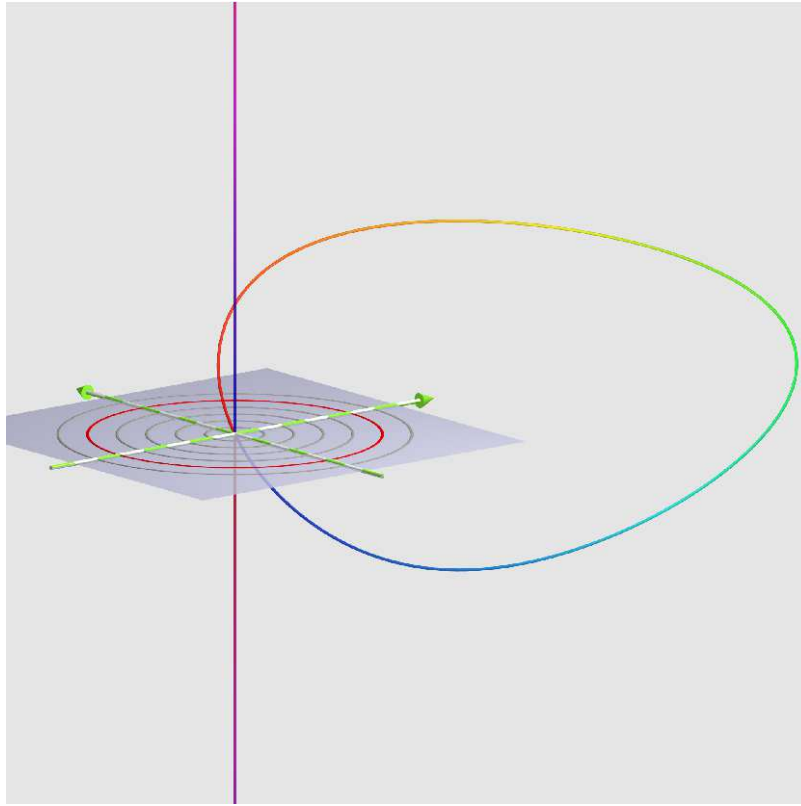}
\end{center}
\caption{
Worldline of an object, which traverses a CTC passing through the origin only once, displayed in the $(t,x,y)$-space. This worldline consists of three distinct parts: (i) the half-line of the $t$-axis with $-\infty<t<0$, (ii) the CTC passing through the origin, and (iii) the half-line of the $t$-axis with $0<t<\infty$. Before and after its time travel, the object thus remains at rest at the origin. The colour change along the worldline indicates the proper time $\tau_\orm$ of the object, where we assume that $\tau_\orm=0$ for $t=0$. Moreover, we have omitted the $t$-axis in order to not occlude the straight part of the worldline. Note that the units are chosen as in figure~\ref{fig:lightconediagram}.}
\label{fig:worldline1}
\end{figure}

\enlargethispage*{2ex}
In order to avoid multiple existence of the travelling object at the same space-time point, we restrict the time travel of the object to one round-trip only. Before the object starts, and after it finishes its round trip, it remains at rest at the origin~$r=0$. In figure~\ref{fig:worldline1} the worldline of such a trip is presented in the $(t,x,y)$-subspace. A remarkable feature is the representation of the object's proper time $\tau_\orm$ along the worldline by colours. The straight part of the worldline reveals that the object is always at rest close to the observer's (slightly displaced) position. At~$t=0$ the object starts to move along the CTC that passes through the origin. After it returns to the latter, the object stops moving in space and remains at rest at the origin. Yet for the observer it seems that the object just flips its proper time at $t=0$ which is clearly visible by the colour shift between the two straight branches for positive and negative coordinate time~$t$ of the worldline. 

\newpage
Furthermore, the worldline suggests the mind-boggling effect that for a certain period the object appears twice in the viewable area of the observer. Indeed, for negative coordinate times $t$, the object from the ''future'' reappears after it crosses the G\"odel horizon (blue branch) whereas for positive times $t$ the object disappears behind the G\"odel horizon (orange branch).

\begin{figure}[p]
\centering
\includegraphics[width=0.75\textwidth]{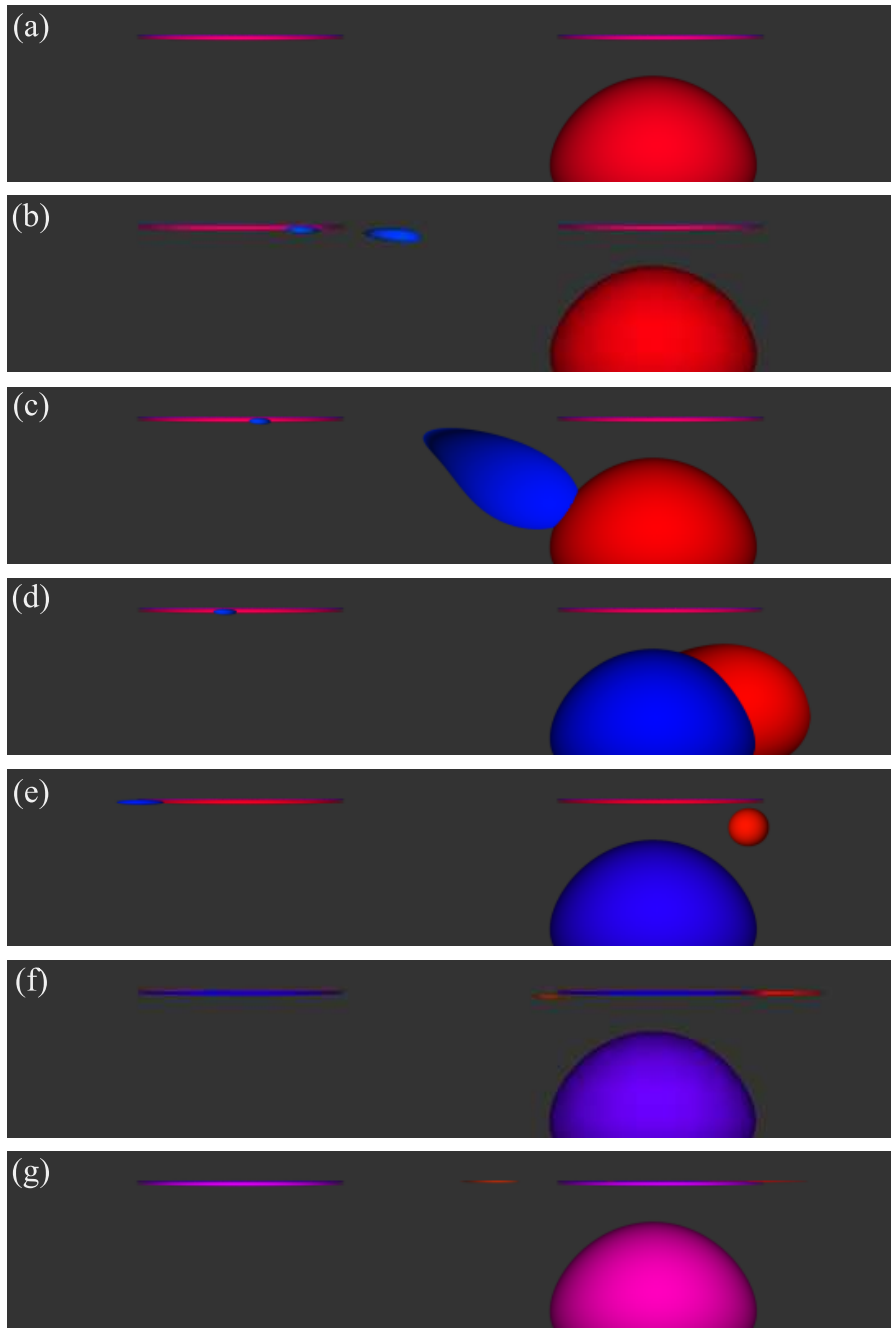}
\caption{Visualizations of a sphere moving along the worldline of figure~\ref{fig:worldline1} represented here for different coordinate times $t$ (see supplementary movie 11, available from \link{stacks.iop.org/NJP/15/013063/mmedia}). The observer looks in the $y$-direction with a horizontal and vertical aperture of $360^\circ$ and $120^\circ$, respectively. We have clipped the top part of the image, since the sphere does not appear there. The sphere has a radius of $0.05$ and the colour of its surface indicates the proper time along the worldline in complete accordance with the colours of figure~\ref{fig:worldline1}. As the worldline suggests, the sphere remains at rest for $t<0$. It is intriguing that another image of that sphere becomes visible showing the sphere already coming back from its journey before it starts at $t=0$. At $t>0$ the ''younger version'' of the sphere moves away from its initial position and disappears behind the optical horizon defined by the G\"odel radius, while the ''older version'
' remains at rest 
close to the observer.}
\label{fig:vis_worldline1}
\end{figure}

In figure~\ref{fig:vis_worldline1} we visualize a sphere with radius $0.05$ travelling along the worldline of figure~\ref{fig:worldline1}. In order to bring out the temporal retardation effects of the visualization once more, the sphere changes its surface colour with elapsing proper time~$\tau_\orm$ in accordance with the colour change along the worldline of figure~\ref{fig:worldline1}. We note that in figure~\ref{fig:worldline1} and~\ref{fig:vis_worldline1} the surface colour displays the proper time of the sphere $\tau_\orm$ at the moment the light was emitted, whereas in figure~\ref{fig:dyn_erde_geo1_eigentime} the colour coding expresses the time difference the light needs to propagate from the globe to the observer.
\newpage

Moreover, to have a better perspective on the sphere, we slightly displace the observer out of the centre by an amount of 0.075 into the negative $x$-direction and $0.075$ into the positive $z$-direction. In this way, the sphere appears slightly below the observer's viewing direction which is parallel to the $(x,y)$-plane in the visualization of this scenario.

The first picture in the visualization (a) shows the sphere (red) at rest at the observer's position. In this phase the sphere's proper time $\tau_\orm$ coincides with the coordinate time~$t$. With elapsing $t$, the same sphere but at a later proper time reappears from behind the G\"odel horizon (b) and is seen as a blue sphere in complete agreement with the colour of the worldline in figure~\ref{fig:worldline1}. As soon as the blue sphere returning from the ''future'' has completed its journey on the CTC at $t=0$ (c), the ``younger'' (red) sphere starts the trip along the CTC (d), from which the blue sphere just arrived. At an even later time the orange sphere vanishes out of sight as it crosses the G\"odel horizon (f). At the end, the original sphere in front of the observer is visible (g) but with a shift in its proper time $\tau_\orm$. 

Other images of the sphere that appear at the observer's horizon are a result of the zeroth order indirect, as well as of the first order direct light rays. For instance, in figure~\ref{fig:vis_worldline1}(a) the left purple image of the sphere stems from the zeroth order indirect light rays, whereas the right purple image of the sphere is due to the first order direct light rays. According to figure~\ref{fig:worldline1}, the purple colours indicate that the corresponding light rays left the sphere at the origin at an earlier proper time~$\tau_\orm$. In the preceding visualizations this phenomenon has already manifested itself in multiple images of the visualized object appearing in the observer's view.

Our visualization of a sphere travelling along a CTC highlights an important feature of time travel in G\"odel's universe: an observer at rest with the field generating matter will never be able to visually perceive the time travelling object entirely on its way along the CTC. For such an observer, the process of time travel remains partially hidden behind his optical horizon.

\section{Summary\label{sec:summary}}
\enlargethispage*{2ex}

Throughout this article visualizations have provided us with a profound insight into optical effects caused by the curvature of G\"odel's universe. Our approach of rendering the presented visualizations is based on the ray tracing method known from standard computer graphics. In order to grasp an understanding of the appearance of the objects, it is essential to determine the trajectories of light rays in the G\"odel space-time.

It turned out that all light rays are of the form of a more or less stretched helix aligned parallel to the $z$-axis. A consequence of this peculiar geometry is that light emitted at a specific point can reach the observer on several paths, depending on the structure of the helix. In order to refer to each ray unambiguously, we have used a classification system which labels each ray according to its number of revolutions and in which direction the light ray is emitted -- either towards the origin or towards the G\"odel horizon.

In our visualizations these different classes of light rays manifest themselves in a series of images of an object. We basically distinguish two clusters of images which arise from the two different directions of emitted light. In each cluster we find an infinite number of images depending on the revolution number of the associated light rays. However, this structure is not always discernible. Depending on the position of the visualized object it can occlude its own light rays. As a result, only very few images can be seen. 

Moreover, the arrangement of images changes significantly when the object moves. In this case temporal retardation effects become essential since the objects change their position while the light propagates on their individual paths to the observer. The result is an obfuscating set of images showing the visualized object for different proper times at different positions.

Another intriguing effect of the helical structure of the null geodesics is that the light rays entail an optical horizon given by the G\"odel radius which limits the viewable area of the observer onto a cylindrical region. However, this horizon does not constitute a physical barrier since the G\"odel universe is a homogeneous space-time. Each object can cross this horizon but then disappears out of sight for an observer located at the origin. Since causality can only be violated by crossing the G\"odel radius, the optical horizon exhibits a protection mechanism for objects from being observed travelling back into the past.

We further took advantage of the isometric transformations of the G\"odel metric to construct appropriate closed time-like curves (CTCs) for our visualizations. The standard circular CTCs in G\"odel's space-time lie completely outside of the viewable area of an observer located close to the origin. By exploiting the intrinsic symmetries of the G\"odel space-time, we converted such standard CTCs into CTCs passing through the origin. This technique enabled us to visually trace an object moving on a CTC which starts from the observer's position.

Peculiar visual effects such as the ones presented for the G\"odel universe are not limited to curved space-times only. There exist analogies~\cite{Schleich84} between the light propagation in curved space-times and light propagation in dielectric media. Indeed, one could imagine to realize an analogue of the G\"odel metric very much in the spirit of the recent progress of the design of meta materials for cloaking. Similarly, these media can be realized by appropriately flowing media~\cite{Leo99, Leo06,Leonhard2010}. For example in \cite{Bru08} Brumfield describes how fluids can mimic the event horizon of a black-hole in terms of a table-top experiment. In case of the G\"odel universe we might achieve the typical helical structure in the vicinity of the observer by a rotating fluid having the same impact as the rotational space-time structure. Nevertheless, CTCs remain a peculiarity of the G\"odel universe, impossible to be realized in a table-top experiment.

\ack\addcontentsline{toc}{section}{{Acknowledgements}}
We thank F.~Grave, T.~M\"uller, H.~Ruder and G.~Wunner for the fruitful collaboration made possible by the financial support of the DFG (German Research Foundation) through the project ``Visualisierung geschlossener zeitartiger Kurven in der Allgemeinen Relativit\"atstheorie''. As members of the QUANTUS project~\cite{Zoe10}, EK and WPS gratefully acknowledge financial support from the German Space Agency DLR with funds provided by the Federal Ministry of Economics and Technology (BMWi) under grant number DLR~50~WM~0837.

\section*{References\addcontentsline{toc}{section}{{References}}}


\begin{thebibliography}{99}
\bibitem{Pfa81} Pfarr J 1981 {\it Gen. Rel. Grav.} {\bf 13} 1073

\bibitem{John90} John F, Morris M S, Novikov I D, Echeverria F, Klinkhammer G, Thorne K~S and Yurtsever U 1990 {\it Physical Review} D {\bf 42} 1915

\bibitem{Echeverria91} Echeverria F, Klinkhammer G and Thorne K S 1991 {\it Physical Review} D {\bf 44} 1077

\bibitem{Haw92} Hawking S~W 1992 {\it Phys. Rev.} D {\bf 46} 603

\bibitem{FefermanVolIII} G\"odel K 1995 Lecture on rotating universes {\it Kurt G\"odel: Collected Works} vol 3 
{\it Unpublished Essays and Lectures} ed S Feferman \etal (Oxford: Oxford University Press) p 261

\bibitem{Rin09} Rindler W 2009 {\it Am. J. Phys.} {\bf 77} 498

\bibitem{Goedel1949} G\"odel K 1949 A remark about the relationship between relativity theory and idealistic
philosophy {\it Albert Einstein: Philosopher-Scientist} ({\it Library of Living Philosophers} vol~7) ed P A Schilpp
(Evanston, IL: Open Court) p 557

\bibitem{Goe49} G\"odel K 1949 {\it Rev. Mod. Phys.} {\bf 21} 447

\bibitem{Costa08} Costa L F O and Herdeiro C A R 2008 {\it Phys. Rev.} D {\bf 78} 024021

\bibitem{Jun06} Jung T 2006 {\it Berichte zur Wissenschaftsgeschichte} {\bf 29} 325

\bibitem{Gam46} Gamow G 1946 {\it Nature} {\bf 158} 549

\bibitem{Hub29} Hubble E 1929 {\it Proc. Natl. Acad. Sci. U.S.A.} {\bf 15} 168

\bibitem{Pfister1995} Barbour J B and Pfister H 1995 {\it Mach’s Principle: From Newton’s Bucket to Quantum Gravity} (Boston: Birkh\"auser)

\bibitem{Ciufolini95} Ciufolini I and Wheeler J A 1995 {\it Gravitation and Inertia} (Princeton: Princeton University Press)

\bibitem{Bondi1997} Bondi H and Samuel J 1997 {\it Phys. Lett.} A {\bf 228} 121

\bibitem{Ker63} Kerr R P 1963 {\it Phys. Rev. Lett.} {\bf 11} 237

\bibitem{Car68} Carter B 1968 {\it Phys. Rev.} {\bf 174} 1559

\bibitem{Got91} Gott~III J R 1991 {\it Phys. Rev. Lett.} {\bf 66} 1126

\bibitem{Bir00} Birmingham D and Sen S. 2000 {\it Phys. Rev. Lett.} {\bf 84} 1074

\bibitem{Fro90} Frolov V P and Novikov I D 1990 {\it Phys. Rev.} D {\bf 42} 1057

\bibitem{Sto37} van~Stockum W J 1937 {\it Proc. Roy. Soc. Edinburgh} {\bf 57} 135

\bibitem{Grave08} Grave F and Buser M 2008 {\it IEEE Transactions on Visualization and Computer Graphics} {\bf 14} 1563

\bibitem{Kaj09} Kajari E, Buser M, Feiler C and Schleich W P 2009 {\it La Rivista del Nuovo Cimento} {\bf 32} 339
\nonum Kajari E, Buser M, Feiler C and Schleich W P 2009 Rotation in relativity and the propagation of light {\it Atom Optics and Space Physics} ({\it Proc. Int. School of Physics ''Enrico Fermi''} Course CLXVIII) ed E Arimondo \etal (Amsterdam: IOS Press / Bologna: SIF) p 45

\bibitem{Fei09} Feiler C, Buser M, Kajari E, Schleich W P, Rasel E M and O'Connell R F 2009 {\it Space Science Reviews} {\bf 148} 123

\bibitem{Kaj04} Kajari E, Walser R, Schleich W P and Delgado A 2004 {\it Gen. Rel. Grav.} {\bf 36} 2289

\bibitem{Delgado02} Delgado A, Schleich W P and S\"ussmann G 2002 {\it New J. Phys.} {\bf 4} 37

\bibitem{Gra09} Grave F, Buser M, M\"uller T, Wunner G and Schleich W P 2009 {\it Phys. Rev.} D {\bf 80} 103002

\newpage
\bibitem{App68} Appel A 1968 Some techniques for shading machine renderings of solids {\it AFIPS '68 (Spring) Proceedings of the April 30--May 2, 1968, spring joint computer conference}  {\bf } (New York: ACM Press) p 37

\bibitem{Schleich84} Schleich W P and Scully M O 1984 General Relativity and Modern Optics {\it New Trends in Atomic Physics (Les Houches 1982)} Session~XXXVI ed G Grynberg and R Stora (Amsterdam: North-Holland) p 995

\bibitem{Leo99} Leonhardt U and Piwnicki P 1999 {\it Phys. Rev.} A {\bf 60} 4301

\bibitem{Leo06} Leonhardt U and Philbin T G 2006 {\it New J. Phys.} {\bf 8} 247

\bibitem{Leonhard2010} Leonhardt U and Philbin T G 2010 {\it Geometry and Light: The Science of Invisibility} (New York: Dover)

\bibitem{Bru08} Brumfiel G 2008 {\it Nature} {\bf 451} 236

\bibitem{Zoe10} van Zoest T {\it et al} 2010 {\it Science} {\bf 328} 1540

\end{thebibliography}
\end{document}